\documentclass[manuscript]{acmart}
\usepackage{xcolor}
\AtBeginDocument{%
  \providecommand\BibTeX{{%
    \normalfont B\kern-0.5em{\scshape i\kern-0.25em b}\kern-0.8em\TeX}}}

\setcopyright{acmcopyright}
\copyrightyear{}
\acmYear{2018}
\acmDOI{}

\acmConference[]{}{}{}
\acmBooktitle{}
\acmPrice{15.00}
\acmISBN{978-1-4503-XXXX-X/18/06}

\begin{document}

\title{Question-Driven Design Process for Explainable AI User Experiences}

\author{Q. Vera Liao}
\affiliation{%
  \institution{IBM Research AI}
  \city{Yorktown Heights}
  \state{New York}
  \country{USA}}
 \email{vera.liao@ibm.com}
 
 \author{Milena Pribić}
\affiliation{%
  \institution{IBM Design}
  \city{Austin}
  \state{Texas}
  \country{USA}}
 \email{mpribic@us.ibm.com}

  \author{Jaesik Han}
\affiliation{%
  \institution{IBM Design}
  \city{New York}
  \state{New York}
  \country{USA}}
 \email{hanj@us.ibm.com}

\author{Sarah Miller}
\affiliation{%
  \institution{IBM Research}
  \city{Cambridge}
  \state{Massachusetts}
  \country{USA}}
 \email{millers@us.ibm.com}

 \author{Daby Sow}
\affiliation{%
  \institution{IBM Research}
  \city{Yorktown Heights}
  \state{New York}
  \country{USA}}
 \email{sowdaby@us.ibm.com}

\renewcommand{\shortauthors}{}

\begin{abstract}
\
A pervasive design issue of AI systems is their explainability--how to provide appropriate information to help users understand the AI. The technical field of explainable AI (XAI) has produced a rich toolbox of techniques. Designers are now tasked with the challenges of how to \textit{select} the most suitable XAI techniques and \textit{translate} them into UX solutions. Informed by our previous work studying design challenges around XAI UX, this work proposes a design process to tackle these challenges. We review our and related prior work to identify requirements that the process should fulfill, and accordingly, propose a \textit{Question-Driven Design Process} that grounds the user needs, choices of XAI techniques, design, and evaluation of XAI UX all in the \textit{user questions}. We provide a mapping guide between prototypical user questions and exemplars of XAI techniques to reframe the technical space of XAI, also serving as boundary objects to support collaboration between designers and AI engineers. We demonstrate it with a use case of designing XAI for healthcare adverse events prediction, and discuss lessons learned for tackling design challenges of AI systems.
\end{abstract}




\maketitle

\section{Introduction}
Artificial intelligence (AI) technologies, especially those using Machine Learning (ML) algorithms, have become ubiquitous even in high-stakes domains such as healthcare, finance, transportation, even criminal justice. How to make AI explainable, that is, providing appropriate information to help users \textit{understand} its functions and decisions, has been considered one of the critical and pervasive design issues for AI systems~\cite{amershi2019guidelines,liao2020questioning}. This understanding is a necessary foundation to serve many user needs such as to improve, contest, develop appropriate trust and better interact with AI, and eventually accomplish the user goal, whether it is to make better decisions or to delegate tasks to AI with confidence. However, at the present time,  explainability features are still not a common presence in consumer AI products. Recent studies, including our own work~\cite{liao2020questioning}, looking into industry practices around explainable AI paint a dim picture where while ``explainability was strongly advised and marketed by higher-ups'' ~\cite{bhatt2020explainable}, product teams are grappling with tremendous challenges to implement the right solutions, and more broadly, a lack of frameworks to decide when, where and how to provide AI explanations.  

The time is ripe to start tackling these challenges and establish a shared understanding and best practices on how to design explainable AI systems. In the past few years, the academic field of explainable AI (XAI) has seen a fruitful surge, thanks to the growing public interest in fair, transparent and responsible AI. Hundreds of, if not more, algorithms that produce various types of explanations have been proposed in the literature~\cite{guidotti2018survey,adadi2018peeking,arrieta2020explainable}. Meanwhile, many open source libraries and toolkits of XAI~\cite{H2o,alibi,AIX,Microsoft} emerged recently, making XAI algorithms originated in the academic community into practitioners' toolbox.

The availability of a toolbox does not guarantee its intended utilization and benefits. First, with a toolbox there is the challenge of selection--how to effectively and efficiently identify a suitable XAI technique for a given application or interaction when there are an abundance of, and rapidly growing, choices. Second, there is the challenge of translation--how to make technical innovation made by academic research effective and usable in real-world sociotechnical systems. This problem is especially prominent for XAI, as the technical XAI field has been criticized for developing algorithms in a vacuum~\cite{miller2019explanation}. Many XAI algorithms were developed with the intention of allowing scrutiny of ML models during model developing process. So the intended users are often ML engineers or researchers, whose abilities to understand AI and reasons to seek explanations may differ remarkably from end users of consumer AI products. To utilize these XAI algorithms in diverse AI products will inevitably require adaption of the algorithms, translation of the algorithmic output, and effectively integrating them in the overall user workflow and experiences. 

Designers and UX researchers of AI products are often burdened with these challenges. Recent studies~\cite{eiband2018bringing,liao2020questioning} looking into design processes of XAI systems break down the design challenges into two parts. One is to answer \textit{what to explain}, i.e., to identify the right explanation content based on product, user, context and interaction specific needs. Second is to address \textit{how to explain}, to create the right presentation format for the explanation, which often requires iterative design and evaluation. These two stages can be seen as corresponding to the challenges of selection and translation to use the toolbox of XAI techniques. 

To tackle the design challenges of XAI also requires tackling general challenges that designers face working with AI systems, which have received growing attention in the human-computer interaction (HCI) and design communities. One challenge is the need to form a "designerly" understanding of what AI can or cannot do to be able to envision possible interactions~\cite{yang2018investigating,girardin2017user,dove2017ux}. This is specially difficult in the context of designing XAI because designers may need to understand not one, but a toolbox of XAI algorithms. This is also especially necessary for XAI because current XAI algorithms cover only a subset of the landscape of explanations in human experiences~\cite{miller2019explanation,mittelstadt2019explaining}. What kind of AI explanations can be provided to users is often constrained by the technical affordance. As our previous study looking into design practices around XAI~\cite{liao2020questioning} found, the critical challenge for XAI UX is ``finding the right pairing to put the ideas of what’s right for the user together with what’s doable given the tools or the algorithms''.

The second challenge is for designers to be able to collaborate effectively with AI engineers (both broaderly defined\footnote{For simplicity, for the rest of the paper, we refer all roles that perform design tasks as ``designer'', which may also include UX researchers, user researchers, etc. We refer all roles that perform engineering tasks on the model as ``AI engineer'', which may include data scientists, AI researchers, etc.})~\cite{yang2018investigating,girardin2017user}. This is a unique problem to AI product teams, as the design goals should be formulated with a shared vision with AI engineers based on the technical viability, and ideally impact the model development and evaluation. However, in reality AI experts are a scarce resources and designers are often found in a challenging situation to collaborate with them due to a lack of common knowledge and shared workflow.  When it comes to designing XAI UX, our previous study found that communication barriers preventing buy-in from data scientists, who are burdened with implementing XAI techniques, are major roadblockers for designers to advocate for AI explainability~\cite{liao2020questioning}.

Recent work suggests that AI may be ``a new and difficult design material''~\cite{dove2017ux,yang2019profiling} that requires developing AI specific user-centered design processes. The process should help designers form a high-level understanding of AI's technical capabilities as they explore the problem and solution space of their design, as well as closer and iterative collaborations with AI engineers instead of treating design as the final deliverable. We further argue that it is necessary to develop AI design-problem specific user centered design processes, which, besides explainability, include fairness, robustness, errors and uncertainty, among others. These are known design issues that are universally present in AI systems, and critical to AI UX that a product team may choose to dedicate their design effort on. Similar to XAI, each of these topics has a vast collection of algorithmic solutions contributed by an an active academic community, and made increasingly accessible by open-source toolkits. When treating algorithmic solutions as ``a toolbox of design materials'', the challenges of selection and translation are likely shared problems. 

In this work, we contribute a novel design process to enable designers and product teams to work with a toolbox of AI algorithms, focusing on the design issue of AI explainability. We start by reviewing related work on XAI and AI design, and identify four requirements that a design process for XAI UX should satisfy. Based on these requirements, and building upon prior work on designing explainable technologies, we propose a \textit{Question-Drive Design Process for XAI}, which grounds identification of user needs and requirements, choices of XAI techniques, design and evaluation of XAI UX solutions, all in the \textit{user questions} to understand the AI. By mapping user questions to exemplars of XAI algorithms, for which we provide a mapping guide, this process also enables the user questions to serve as boundary objects to support consensus building between designers and AI engineers.


Below we first review related work, leading to the requirements that guided our proposal. Then we describe in detail the \textit{Question-Driven Design Process for XAI UX}, and demonstrate it with a case study of designing XAI for healthcare adverse events prediction following this process. Lastly, we discuss considerations and feedback for practitioners to adopt this design process, and broader implications for designing AI UX.

\section{Related Work}

\subsection{Explainable AI}
While the origin of explainable AI could be traced back to earlier work in expert systems~\cite{swartout1983xplain}, the re-surge of this wave of XAI work could be attributed to an increasing popularity of opaque, difficult-to-understand machine learning (ML) models such as those using deep neural networks. The academic research of XAI was accelerated by legal requirements such as EU's General Data Protection Regulation (GDPR) and various government funding programs~\cite{gunning2017explainable}, as well as broad industry awareness for explainable and responsible AI. In short, the field has seen booming efforts in the past five years or so, developing a vast collection of new algorithms. 

While there is no agreed-upon definition or criteria of AI explainability, and there are different related terms used in the literature such as interpretability and transparency, a common goal shared by XAI work is to make AI's decisions or functions \textit{understandable} by people~\cite{guidotti2018survey,miller2019explanation,mittelstadt2019explaining,arrieta2020explainable,adadi2018peeking,arya2019one,lipton2018mythos}. This understanding can then support various goals for people to seek explanations, such as to assess AI capability, trustworthiness and fairness, to diagnose and improve the model, to better control or work with AI, and to discover new knowledge~\cite{liao2020questioning,adadi2018peeking,miller2019explanation}. This wave of XAI work has focused dominantly on ML models, although explainbility of other types of AI systems such as planning~\cite{chakraborti2020emerging}, multi-agent systems~\cite{rosenfeld2019explainability}, etc. are receiving increasing attention. Following this trend, this paper will primarily focus on designing explianable ML systems.

A complete overview of the landscape of XAI algorithms is beyond the scope of this paper. Multiple recent papers surveyed this field and converged on a few important dimensions to classify and characterize XAI algorithms~\cite{guidotti2018survey,arya2019one,arrieta2020explainable}. One is the differentiation between directly explainable models and opaque models, the latter of which often require using separate algorithms to generate \textit{post-hoc} explanations. In terms of the scope of explanation, XAI work differentiates between global explanations, which provide a general understanding of how the model works, and local explanations, which focus on explaining how the model makes a prediction for a particular instance. Guidotti et al.~\cite{guidotti2018survey} further differentiates a third category: model inspection, which focuses on explaining a specific property of a model such as how changes of a feature impact the model prediction. More fine-grained classification methods consider the formats of the explanation, which could range from less than 10~\cite{guidotti2018survey} to more than 40 types~\cite{arrieta2020explainable}, each of which could be implemented by multiple algorithms that differ in computational properties. 

In short, the technical field of XAI have produced hundreds of algorithms that produce various forms of explanation. Recently, open-source toolkits such as IBM AIX 360~\cite{AIX}, Microsoft InterpretML~\cite{Microsoft}, H2o.ai Machine Learning Interpretability~\cite{H2o} are making popular XAI algorithms (e.g.~\cite{ribeiro2016should,ribeiro2018anchors,dhurandhar2018explanations,lundberg2017unified}) increasingly accessible to practitioners. It has become necessary to develop methods that could help practitioners identify XAI algorithms suitable for their purposes. Such methods should be actionable for practitioners to follow, and scalable as the field is still rapidly advancing.

\subsection{HCI  and user-centered perspectives of XAI}

The HCI community have a long-standing interest in algorithmic transparency and making computing systems explainable. Most notably, a 2018 paper by Abdul et al.~\cite{abdul2018trends}  conducted a literature analysis of HCI research on explainable systems, ranging from expert systems, recommenders, context-aware technologies to more recent ML systems. The authors identified opportunities for HCI research agenda to focus on improving the rigor, by testing effectiveness of explanation interfaces in real-world systems, and improving the usability, by drawing from rich body of cognitive science, design and HCI knowledge on how people consume explanations. Similar views have emerged in the AI community pushing towards a human-centered instead of algorithmic centered XAI field, calling out gaps between XAI algorithmic output and properties of explanations sought by people~\cite{miller2019explanation}, and the necessity to produce and evaluate AI explanation by involving the intended user group in the intended usage context~\cite{doshi2017towards}.

HCI researchers have begun designing explainable AI systems by utilizing various XAI algorithms and making them usable. Some studied explainable interfaces to help model developers diagnose and improve ML models~\cite{kaur2020interpreting,hohman2019gamut}.  Others studied explainability features to help decision-makers such as physicians~\cite{xie2020chexplain,wang2019designing} and recruiters~\cite{cheng2019explaining} to better understand AI recommendations and make informed decisions. It is common for these systems to include multiple XAI features, as users often demand different kinds of explanations at different usage points. Novel usage of AI explanations have also been explored. For example, Lai et al.~\cite{lai2020chicago} leveraged XAI algorithms to generate rationales of toxicity detection as tutorials to train people to perform content moderation. Ghai et al. ~\cite{ghai2021explainable} proposed explainable active learning, which utilizes AI explanations as interfaces for annotators to provide feedback to train models. Another active research area focuses on rigorously evaluating the effectiveness of XAI techniques~\cite{dodge2019explaining,alqaraawi2020evaluating,yang2020visual,cai2019effects,buccinca2020proxy}. For example, Dodge et al.~\cite{dodge2019explaining} compared the effectiveness of multiple popular forms of XAI techniques in supporting people's fairness judgment of ML models, and found that there is no one-fits-all solution. 


These HCI studies have generated valuable insights for tackling the challenges of selection and translation of the toolbox of XAI algorithms. They often suggest domain or user group specific guidelines on what kind of XAI technique works effectively, and what interface features should be provided. In a parallel effort, recent work proposed several taxonomies of prototypical roles of XAI consumers~\cite{arrieta2020explainable,hind2019explaining,tomsett2018interpretable}, as a starting point to provide role-based guidelines to choose between XAI techniques. Typical user roles include model builders, decision-makers, people being impacted by AI, business owners, regulatory bodies, etc.

\begin{figure*}
  \centering
  \includegraphics[width=1\columnwidth]{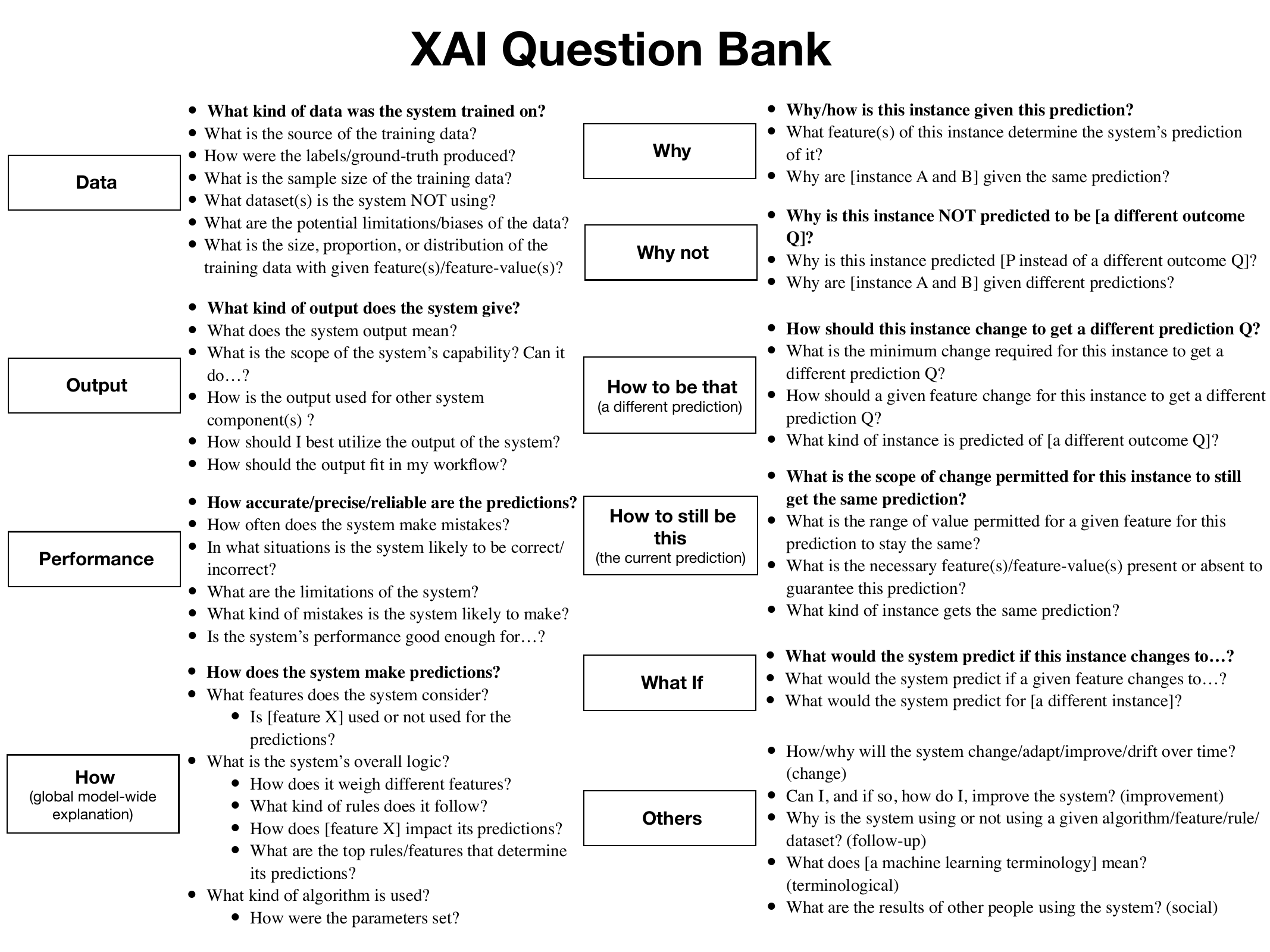}
   \vspace{-2em}
  \caption{XAI Question Bank for explaining supervised machine learning systems we developed in~\cite{liao2020questioning}, with minor updates }~\label{fig:question}
   \vspace{-2em}
\end{figure*}

While these efforts can be seen as producing top-down guidelines, recent studies begun to explore bottom-up, user-centered approaches to designing XAI UX~\cite{liao2020questioning,eiband2018bringing,wolf2019explainability}. This paper is built upon our previous work~\cite{liao2020questioning} studying design challenges around XAI by interviewing 20 designers working across 16 AI products. We identified a key challenges for creating XAI UX is the variability of users' explainability needs, i.e. types of explanation sought. Based on prior HCI work on ``intelligibility types"~\cite{lim2009and} and social science literature on explanatory relevance~\cite{hilton1990conversational} that defines different types of explanation, we argued to represent user needs for explainability by the type of question the user asks to understand the AI, such as \textit{Why}, \textit{What if}, \textit{How}, etc. Our study found that many factors could vary the type of question a user asks, including their motivation to seek explanation, points in a user journey, model type, decision contexts, among others. This observation highlights that guideline-based approaches may not be granular enough to allow system, user, and interaction specific selection and design of XAI techniques, and thus a user-centered approach to ground the design in a nuanced understanding of users and the usage contexts is often necessary. Furthermore, in that study we ``crowdsourced'' common user questions from the designers and created an XAI Question Bank, shown in Figure~\ref{fig:question} (with minor updates from~\cite{liao2020questioning}). The questions are organized by a taxonomy of 9 categories, representing prototypical questions people have for different types of AI explanation, focusing on ML systems. Following this effort, in this work we will propose a user-centered design process for XAI that starts by identifying the types of explanation needed by eliciting user questions. The proposed design process will also use the XAI Question Bank (Figure~\ref{fig:question}) as an auxiliary tool.

\subsection{What's unique about designing UX of AI }
Our proposed method is also motivated by recent work looking into the unique challenges of designing user experiences for AI systems. A common challenge is the need for designers to understand ML capabilities~\cite{dove2017ux,yang2019profiling,yang2018investigating}. Interestingly, prior work reveals that the required is often not technical knowledge of ML but a ``designerly'' understanding of what a ML technique can or cannot do, and what user values it can deliver. According to Yang et al.~\cite{yang2018investigating}, designers often rely on abstraction of ML capabilities and exemplers of existing AI systems to establish such a designerly understanding. A related area of research that could support a designerly understanding of and formulating design goals for AI aims to develop design guidelines specific to AI systems. Notable examples include Amerishi et al.'s heuristic guidelines for human-AI interaction~\cite{amershi2019guidelines}, and Madaio et al.'s checklist for AI fairness ~\cite{madaio2020co}.

Another area of challenges arise from the unique design process of AI systems. This process often takes longer than non-AI products and should ideally involve designers and AI engineers in a tight collaboration from concept development, co-evolving of design and technical development, and continued refinement after deployment~\cite{yang2018investigating}. However, there is currently no formal requirement or best practices for designers and AI engineers to collaborate. Designers' interactions with AI engineers are often ad-hoc, sporadic and no shortage of communication challenges. By reflecting on the lifecyle of how data scientists build ML models and designers' work process, Giradin and Lathia~\cite{girardin2017user} identified a series of touch points for designers to work with data scientists to produce meaningful UX for AI powered systems, including co-creating a tangible vision of experience with priorities, goals and scope, assessing the assumption of algorithms with insights from user research, articulating solutions and understanding the limitations from both sides, and specifying successful metrics of the user experience that may impact choices of data and algorithms. 

This challenge for designers to work with AI engineers was also foregrounded in our previous study on design challenges around XAI~\cite{liao2020questioning}. Because providing explanability features often requires AI engineers to implement either additional algorithms or functions to generate information about the model, communication barriers and failure to reach a shared design vision often impede buy-in from AI engineers, especially when explainability is at odds with other product goals such as time pressure to deliver minimum viable products (MVPs).

Other design challenges of AI UX discussed in the literature include a shift towards data-driven design culture ~\cite{kun2019creative,bogers2016connected}, challenges to prototype and engage in quick design iterations~ \cite{yang2018investigating,van2018prototyping,sun2020developing,begel2020lessons}, and the needs to engage stakeholders to align the values of AI system~\cite{lee2020co,yu2020keeping}. Our work will primarily tackle supporting a ``designerly'' understanding on the technical space of XAI, and designer-AI engineer collaboration, as discussed in the requirements below.

\section{Question-Driven Design for Explainable AI User Experiences}
We first discuss the requirements that guided our proposal, and the method we follow to develop \title{Question-Drive Design Process for XAI UX}. Then we provide an overview of the design process, and describe a use case in which we introduced this design process to a team working on an AI system that provides healthcare adverse event predictions. We elaborate on each stage of the design process, together with the design activities we carried out working with the use case and their outcomes. 

\subsection{Requirements and Development Method of the Design Process}
Based on related work reviewed above, we set out to propose a design process for creating XAI UX with the following requirements: 

\begin{itemize}
    \item \textbf{R1: }\textit{Following a user-centered approach to identify user, context and interaction specific needs for AI explainability}.  Following our prior work~\cite{liao2020questioning}, we ground user needs, specifically what type of explanation is sought, in the type of \textit{question} that a user asks to understand the AI.
    
    \item \textbf{R2: } \textit{Enabling a designerly understanding on the affordance of different XAI techniques available}. We do so by mapping prototypical user questions to candidate XAI technical solutions to reframe the techincal space, focusing on ones that are available in popular open-source toolkits, and providing exemplars of explanation they could generate (in verbal descriptions or visual examples). In other words, this mapping explicitly links existing XAI techniques with values they could provide to users, as grounded in the type of user question they could answer. 
    
    \item \textbf{R3: } \textit{Supporting collaborative problem solving between designers and AI engineers to identify XAI UX solutions that are both technically viable and satisfying user needs}. Specifically, the user questions can be used as \textit{boundary objects} to facilitate the communication and consensus building, in which designers can ground their understanding on user requirements for asking a question, and AI engineers can ground their understanding on the technical details, supported by the mapping guide mentioned above.
    
    \item \textbf{R4: } \textit{Supporting an end-to-end, iterative design process by articulating the design goals or success metrics of XAI to guide continued, in-parallel refinement of design and technical solutions}. 
\end{itemize}{}

With these requirements in mind, we developed Question-Driven Design Process for XAI first through a series of ideation workshops with a group of 11 volunteers, consisting of designers, data scientists and XAI researchers from multiple AI product lines of a large international technology company. We then presented the initial idea of the design process using a slideshow to an additional 10 designers, HCI researchers and data scientists for feedback, and iterated on the content. 

So far the design process has been carried out in two use cases for validation and refinement. One is an end-to-end design process for a team to develop an explainable AI system that provides healthcare adverse event prediction to help healthcare providers identify high-risk patients and improve their care management. We will use this use case to illustrate the design process in the following sections. Another use case is an ongoing project to develop explainable AI solutions for crime analytics. So far the team has adopted the first half of the design process to understand user needs and requirements for XAI. It provides a more complex use case where users are not just receivers of AI predictions but also need to configure their own models. This use case helped us refine the instruction of the design process to make it flexible for different teams' needs, which we will reflect on in later sections.

\subsection{Design Process Overview}

\begin{figure*}
  \centering
  \includegraphics[width=1\columnwidth]{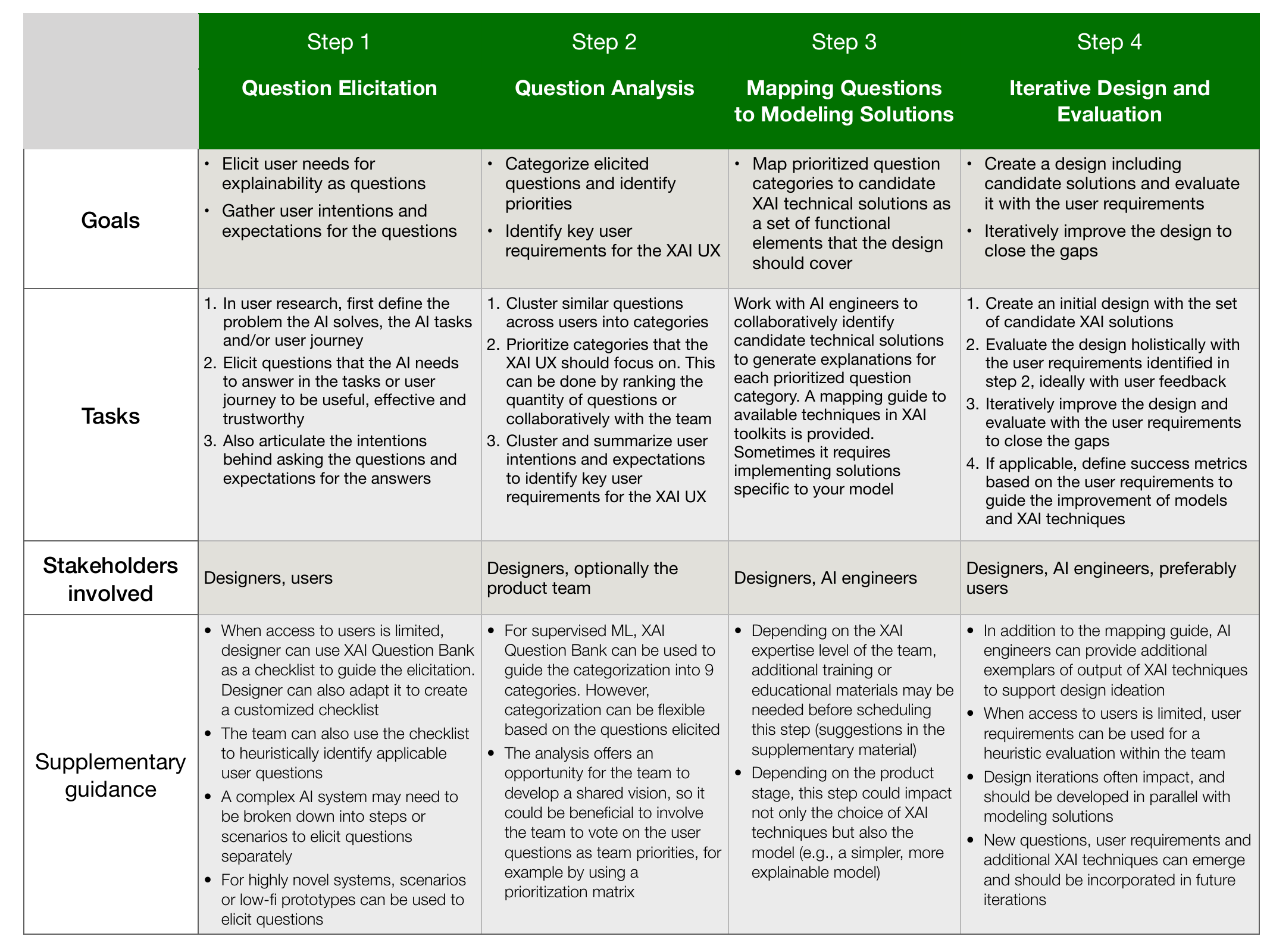}
   \vspace{-2em}
  \caption{Overview of Question-Driven Design Process for XAI UX }~\label{fig:process}
   \vspace{-2em}
\end{figure*}

Figure~\ref{fig:process} provides an overview of the design process, consisting of four steps: question elicitation, question analysis, mapping questions to modeling solutions, and iterative design and evaluation. In the forth row we suggest the stakeholders that should be involved in each step. As we sought feedback for the design process from designers and AI enginers, we became aware that AI product teams differ significantly in their ability to access users, availability of AI experts, designers' as well as the team's general AI expertise, and the nature of the AI technology. Therefore, in the last row we suggest possible workarounds and additional resources if a product team face difficulty executing a step, which we will discuss as \textit{supplementary guidance} in later sections.

We also provide in the supplementary material of this paper a slideshow for the design process, with which we are introducing this design process to product teams by hosting workshops and as an introductory handout. In addition to the detailed description of the design process and the use case to be described, the slideshow also includes an overview of what XAI is and more detailed exemplars of XAI techniques with visual illustrations. In the Discussions section, we will reflect on the lessons learned on how to introduce a new design process to product teams. 

The first two steps, question elicitation and analysis, are to enable designers and the team to understand the user needs and requirements for the XAI UX. The last two steps, mapping modeling solution and iterative design and evaluation, are for designers and AI engineers to engage in collaborative problem solving for technology and UX co-development based on user needs and requirements. 

Based on \textbf{R2} identified above, the collaborative problem-solving could be accelerated by supporting a designerly understanding on the affordance of different XAI techniques. We do so by mapping user question to candidate solutions of XAI techniques, with example descriptions of their output. In practice, we observed that currently AI engineers in product teams are often not familiar with various XAI algorithms. So we also link these candidate solutions to readily usable toolkits or code libraries to support AI engineers' technical understanding of XAI solutions and accelerate the implementation, which could help fulfilling \textbf{R3}. 

\begin{figure*}
  \centering
  \includegraphics[width=1\columnwidth]{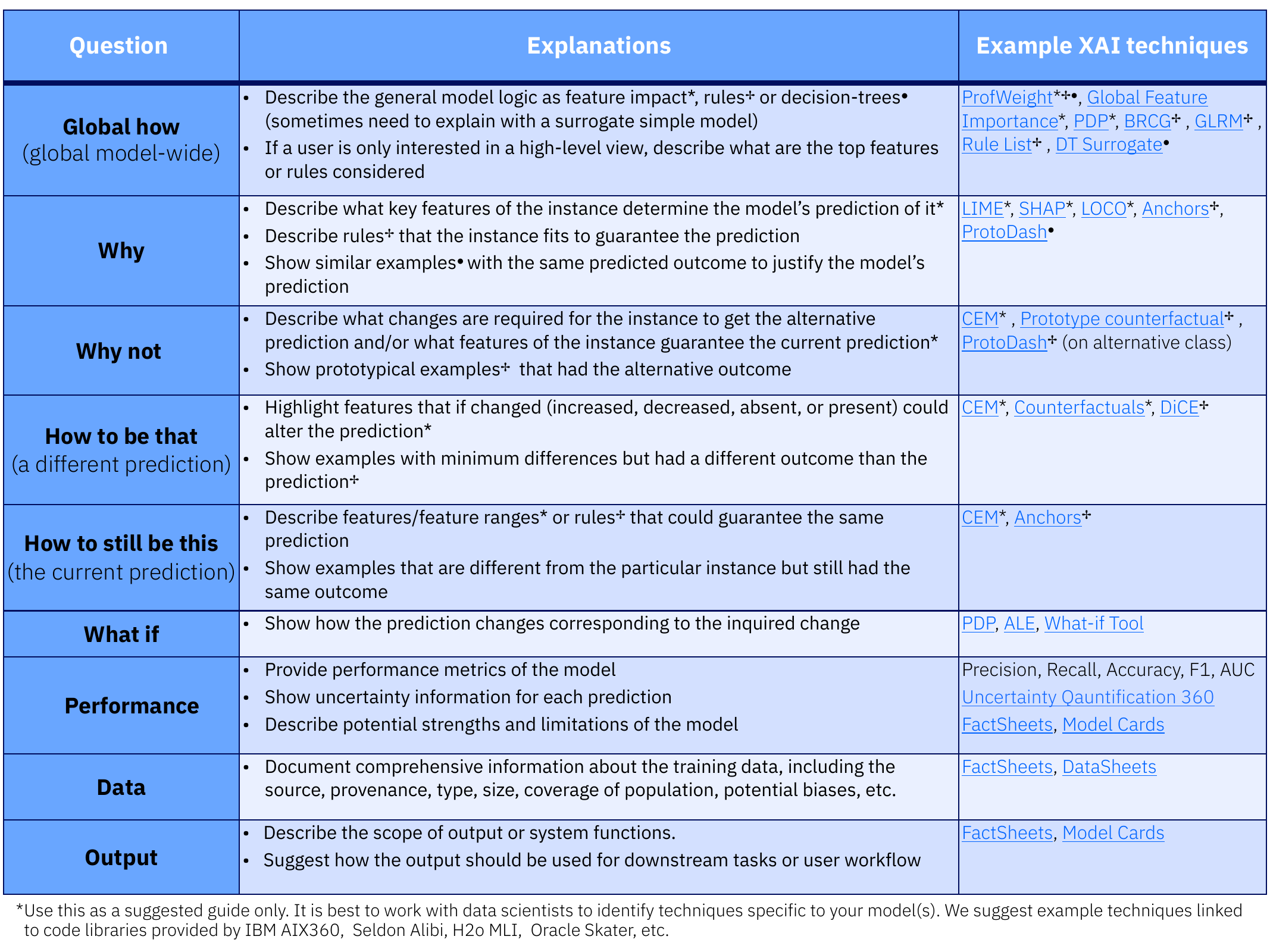}
   \vspace{-2em}
  \caption{Mapping guide between question categories and candidate XAI techniques, with exemplary descriptions of explanations provided by these techniques and links to code libraries }~\label{fig:mapping}
   \vspace{-2em}
\end{figure*}

In Figure~\ref{fig:mapping}, we provide a suggested mapping guide between user questions and descriptions of candidate technical solutions (in the slideshow we also provide visual exemplars), as well as links to code libraries. These questions are based on the categories in the XAI Question Bank~\cite{liao2020questioning} (Figure~\ref{fig:question}). Most links are for post-hoc XAI algorithms, i.e., they can be used to generate explanations for complex models that are not directly explainable. When using simpler models such as regression or decision tree, it is possible to generate such explanations directly based on model internals using standard data science packages, as we elaborated in the slideshow. The last three categories are considered "model facts" to describe the model's performance, training data or output. The corresponding links (FactsSheets~\cite{factsheets}, Model Cards~\cite{model_card} and Datasheets~\cite{gebru2018datasheets}) provide exemplars of industry standards for model documentation. This mapping was built by reviewing various survey papers of XAI techniques~\cite{guidotti2018survey,adadi2018peeking,arrieta2020explainable}, and iterated on with feedback from 5 data scientists and XAI researchers. In particular, we focused on suggesting XAI techniques that are available in current XAI toolkits.

We emphasize that this mapping guide is suggestion only and focuses on explaining standard supervised ML technologies, in which users receive predictions or recommendations from AI to assist their decision-making. In practice, it is possible that an AI system could invoke new categories of user question, or a product team may choose to categorize user questions in different ways. It is important for designers and AI engineers to work together to identify model specific solutions. Sometimes the solution may lie outside the scope of this mapping guide but requires generating information about a specific AI model's parameters or internals, for which the team could use standard data science frameworks or implement a new technique.  

\subsection{Use Case Overview}
In 2020, a team in the technology company were working with a large hospital to develop an AI system that predicts patients' risk of healthcare adverse events, for example unplanned hospitalization due to an urgent and unexpected condition. The model prediction is made based on a patient's profile such as medical history and social-demographic information. The intended users are healthcare providers, who could use the AI's predictions to identify potential high-risk patients to better manage their healthcare plans. It was recognized that explainability, to enable healthcare providers to understand the AI and its predictions, would be critical to the user experience. We joined the project team and introduced the question-driven design process, and worked with the team to carry out the process.

The multi-disciplinary team consisted of designers, UX researchers, data scientists, software engineers, XAI researchers, and a project manager. The team conducted user research with 9 participants (doctors and research nurses). 4 of them are sponsored users from the company and 5 are recruited in the hospital. The prediction model is a Deep Learning model based on custom recurrent neural network (RNN) architectures and trained mainly with medicare claim data. The details of the model and data collected in the hospital are considered proprietary and will not be disclosed in this paper (we will exclude quotes from participants recruited in the hospital). Instead, we will focus on presenting the activities and outcomes by following the question-driven design process. We will not delve into the details of the user research as our goal is to illustrate the design process rather than presenting empirical insights from the use case.

\subsection{Step 1: Question Elicitation}
The first step \textit{Question Elicitation} aims to gather empirical data to understand user requirements for the XAI UX. This step mainly involves designers and targeted users participating in the user research. While it can be part of a broader formative user research using interviews, focus groups, open-ended surveys, etc, we suggest to center the inquiry for explainability around what \textit{questions} users have for understanding the AI. The notion of representing user needs for explainability, or the type of explanation they seek, as questions can be found in HCI literature ~\cite{lim2009and,liao2020questioning} as well as social sciences literature~\cite{hilton1990conversational}.  

The empirical data gathered in this step will be used to fulfill \textbf{R1}, to identify what types of explanations are needed for a specific user group, a usage context and or an interaction, and \textbf{R4}, to help articulating the design goals of the XAI UX. Therefore, we suggest to engage in two parts of user inquiry. First is to ask participants to come up with questions they have for the AI. A prompt we found useful is ``\textit{What questions does the AI need to answer for it to be useful, effective and trustworthy?}'' Second is to follow up and ask participants to articulate the intentions behind asking a question, and their expectations for answers to the question. Prompts we found useful include ``\textit{Why would you ask this question}'' and ``\textit{What would a good answer to this question look like?}''

The question inquiry should happen after defining the problem space and the AI tasks, so that participants can effectively engage in such a reflection. The task definition can either be done with the participants as part of the user research, or provided directly to scaffold the discussion. 

\subsubsection{Activities in the use case}
As we conducted formative user research for the AI system for healthcare adverse event prediction, we engaged in question elicitation as part of the user interviews with 9 participants (doctors and research nurses). The whole interviews lasted 45-60 minutes each. The first half focused on a broader understanding of the task of preventing healthcare adverse events, asking questions such as ``\textit{How do you usually assess patients' risks}'' and ``\textit{what do you do to prevent unplanned hospitalization?}'' Then we asked participants to envision ``\textit{How can an AI system help with adverse events risk prediction and prevention}?'' Depending on participants' experience level with AI, we selectively provided high-level descriptions of AI prediction capabilities and guided the discussion towards envisioning concrete tasks the AI can perform. Participants converged on the vision of an AI system assessing patients' risks or alerting high-risk patients for them to pay attention to, and helping them identify appropriate interventions. At this point, we begun the question elicitation by asking:
\begin{itemize}
\item[--] \textit{Imagine using such an AI system, what do you want to understand about it? What questions would you ask?}
\item[--] \textit{What questions would you want the AI to answer when you look at its risk prediction for a patient?}
\item[--] \textit{Are there any other questions the system should be able to answer for you to trust or feel comfortable using it?}
\end{itemize}

After participates articulated their questions, we followed up by probing for their intentions to ask these questions, and their expectations for the answers whenever there were non-obvious ones, by asking:
\begin{itemize}
\item[--] \textit{Why would you be interested in asking these questions?}
\item[--] \textit{What would be a good answer look like? What should the AI tell you?}
\end{itemize}

All interviews were video recorded and transcribed. Below we focus on the part of data around elicited questions. Sometimes user questions appeared in other parts of the interviews without the elicitation prompts. We also included these questions and discussions around them.

\subsubsection{Outcomes in the use case}
On average, we elicited 6 questions from each participant, with a maximum of 12 questions and minimum of 3 questions. We observed that participants' experience with AI impacted the quantity and breadth of questions they formulated. For example, P1 is a sponsored user recruited in the company, who has had extensive experience working with AI systems. She came up with 12 questions that were clustered in three areas. The first area is about understanding the AI's risk prediction for a patient:
\begin{itemize}
\item[--] \textit{What is the risk the patient has and what can the outcomes be?}
\item[--] \textit{Why this risk?}
\item[--] \textit{Is there a risk because...[they have not taken the drug as prescribed]?}
\item[--] \textit{What are the main risk factors for this person?}
\item[--] \textit{What should I be worried about this patient?}
\end{itemize}

Based on the taxonomy of XAI Question Bank, the first one is an \textit{Output} question, and the last four are \textit{Why} questions. Without us asking, P1 proactively elaborated on the intentions and expectations for asking these questions: ``\textit{it would be helpful if the AI could help me really understand the patient...is there a chronic condition, the reasons they were being admitted...should I put them in a risk of mobility or even mortality }''; and: ``\textit{...understanding why this risk, ideally in a way that I can either address, or acknowledge that, or make some kind of change}''.

Then she moved on to the second area that directly asked about changes she could make to reduce the patient's risk:

\begin{itemize}
\item[--] \textit{Is there some kind of adjustment or accommodation I can make for the patient's care?}
\item[--] \textit{For patients with the same risk profile, what have been done for the success stories?}
\item[--] \textit{What happened to the unsuccessful stories [of patients with the same risk profile] ?}
\end{itemize}

The first two are \textit{How to be that} (less risk) questions. And the last one is a \textit{How to still be this} question. They were all concerned about seeking appropriate interventions, as P1 followed up to explain the intention: ``\textit{Keep[ing] myself honest about what I am going to do [with this patient], whether it is aligned with what has been practiced... guidelines are for broader population, but they are not personalized enough. So it would be better to know what would specifically work for patients like mine.}''

After we further probed on what questions would make her feel comfortable using the system, she asked about the algorithm (\textit{Global how}) and \textit{Data}.

\begin{itemize}
\item[--] \textit{What is the model or algorithm that is generating the predictions?}

\item[--] \textit{Why is this model chosen?}
\item[--] \textit{What are the demographics of the population in the training data?}
\item[--] \textit{What medical conditions do patients in the training data have?}
\end{itemize}

She followed up with rich discussions about her intention for asking these question, as mainly to understand the AI's limitations to decide when to confidently rely on it, and the expectation for the explanation, to be both concise and actionable: ``\textit{..a little bit of background, or insights on what the AI is, what does it applies to. Just a little bit! ... It might be too much for some people, but it would be nice to have a cheatsheet of the methodology. Sometimes people just say we use deep learning or we use logistical progression or whatever. But why? I just want to know, basically, if I use this version, does it mean I might miss out a little bit of this, or does it take the bigger picture thing... As my own due diligence, understanding why it was chosen, because at the end of the day we are responsible for what this does. Especially when [its prediction] doesn't click  with my clinical experience I will have to actually review it and just be like, so this is why so there could be a chance that there's a little off. When it comes to this kind of scenarios, maybe I should lean more of my own experience}''

In summary, a total of 54 questions, and 37 excerpts of discussion on the intention and expectation for asking the questions were collected in Step 1, and will be used for analysis in Step 2. 

\subsubsection{Supplementary guidance} In the use case, we were able to recruit 9 participants. A common feedback we got is that some projects may have limited access to interviewing users, especially in an agile environment or in the initial concept development stage, so they may not be able to gather a sufficient amount of questions to ensure a thorough understanding of user needs. One workaround, we suggest, is to use XAI Question Bank (Figure~\ref{fig:question}) as a checklist, asking users to go through them and identify applicable questions. This approach could enable a more comprehensive coverage of the space of XAI with a small number of participants. It is even possible for the the product team to use the checklist to heuristically identify applicable user questions, which could be useful in the early stage of concept development. The XAI Question Bank~\cite{liao2020questioning} has 50 questions and intentionally include common variations. Designers can also start with selecting from or tailoring these questions to create a customized checklist.

Another flexible decision point that can be tailored to a design project is the context of question elicitation--how well-defined the AI tasks should be, and how to define them. In our use case, we asked participants to define the AI tasks, because we were essentially introducing an AI system to assist an existing well-defined problem in their job. The interview protocol, by asking them to first reflect on current practices and the problem space, effectively facilitated them to define concrete and mostly consistent AI tasks. However, this is likely more challenging if the intended AI tasks are complex or less familiar. If the AI tasks are ill-defined or vary significantly across participants, the question elicitation may generate less value. We suggest several possible workarounds. First, for a complex AI system, it may need to be broken down into steps or scenarios to elicit questions separately. Providing a user journey map or user scenarios can be useful. Second, for highly novel systems that are hard to come up with questions for, user scenarios or even low-fi prototypes can be used to scaffold the elicitation.

\subsection{Step 2: Question Analysis}
The second step \textit{Question Analysis} is to analyze the data collected in Step 1, including the questions and content discussing the intentions and expectations. The goal is to establish an understanding on the user needs and requirements to guide the design and evaluation of XAI UX, both in terms of the types of explanation to provide (\textbf{R1}), and key user requirements as design goals (\textbf{R4}). This step can be completed by designer(s) who then present the insights to the team, or serve as an opportunity to establish a shared team vision by involving the whole team in analyzing the questions and collaboratively identifying priorities (\textbf{R3}).

The first step is to cluster similar questions across participants to arrive at main categories of user questions. For standard supervised ML systems, our XAI Question Bank~\cite{liao2020questioning} (Figure~\ref{fig:question}) suggests main categories and variations of question that could help guide the categorization. Depending on the nature of the AI system and user needs, one may identify new categories of questions that are not presented in XAI Question Bank. 

After the categories are established, the task is to identify which categories should be priorities for the XAI UX to focus on. This prioritization task can be approached in multiple ways. One is to simply count the number of questions under each category, thus the occurrences of such needs across different users, and prioritize those asked most frequently. Or a team can review the questions and decide on the priorities collaboratively

Another area of the analysis is to identify main themes from participants' comments on their intentions and expectations for asking the questions, as key user requirements for XAI UX. This can be done by following standard qualitative analysis approach, by first summarizing or tagging individual excerpts and then clustering similar ones to identify emerging themes. These user requirements will then be used to guide the evaluation of the XAI UX in later steps. We highlight that there is often not a one-to-one mapping between questions and intentions. Participants may ask different questions for the same underlying reason, or similar questions but for different reasons. So we recommend to analyze and cluster the user requirements separately. It may also be necessary to refer to other parts of the interview, for example discussions about the user background or workflow, to better interpret the comments about why or what a user needs by asking a question.

\subsubsection{Activities in the use case}
Two authors of the paper worked with designers to perform the above-mentioned analyses and presented the insights to the team. Since adverse events prediction is a standard supervised ML regression task, we found the questions largely aligned with the taxonomy provided in the XAI Question Bank. However, a challenge we noticed is the need to translate questions to the general forms appeared in the XAI Question Bank. One reason is that the healthcare domain has its own familiar jargon, for example using the terms ``sensitivity and specificity" instead of standard ML performance metrics. Another reason is that participants often asked questions based on their workflow, in a more specific way than the form appeared in the XAI Question Bank. For example, for a \textit{Why} question about understanding the risk factors (features) of a patient, participants asked ``\textit{What comobility factors does this patient have [to make him high-risk]}'', or ``\textit{What's the reason? Is it something in his medical history, social environment, fitness level, etc.?}''

For question prioritization, we chose to simply rank the categories by the number of elicited questions in them. For the user requirements analysis, we conducted standard qualitative analysis by first tagging individual excerpts, and then cluster them to identify emerging themes.

\subsubsection{Outcomes in the use case}
Our analyses revealed four main categories of questions that should be the priorities for the XAI UX to address:
\begin{itemize}
    \item \textit{Why}: All participants asked \textit{Why} questions to understand the reasons behind a prediction, or risk factors of a patient. 
    \item \textit{How to be that:} 6 out of 9 participants asked how or what they could do to reduce the risk or prevent adverse events. This type of question was sometimes asked in a more implicit form. For example P1, as we showed above, asked about what was successfully done to other similar patients.
    \item \textit{Data}: 5 out of 9 participants asked about the training data. This was somewhat surprising given that data was not an explicitly mentioned element when defining the AI tasks. It suggests that healthcare providers have a general understanding of ML models and are concerned that misalignment between training data and their own patients could hamper the system's reliability.
    \item \textit{Performance}: 4 participants asked about the AI's performance. Interestingly, none of the questions mentioned standard performance metrics, but are concerned about the AI's limitations, for example, ``\textit{on what patient population might it work worse}'', or ``\textit{how well does it work on non-obvious cases or patients with certain conditions?}'' 
\end{itemize}{}


Our analysis on the content about participants' intentions for asking the questions and expectation for the answer yielded the following five user requirements for the XAI UX. For a user, the XAI UX should:

\begin{itemize}
    \item \textit{UR1}: Help me discover (otherwise non-obvious) information to better understand patients
    \item \textit{UR2}: Help me determine the most actionable next steps for the management of patients
    \item \textit{UR3}: Help me decide whether I should spend more time on further evaluation or management of a patient
    \item \textit{UR4}: Allow me to assess the reliability of a particular AI prediction
    \item \textit{UR5}: Increase my confidence to use the AI system
\end{itemize}{}

Again, we emphasize that it was not a one-to-one mapping between question types and user requirements. For example, when asking a \textit{Why} question, participants' reasons ranged between \textit{UR1}, \textit{UR2}, \textit{UR3} and \textit{UR4}. In the following steps, we will evaluate the design holistically with the set of user requirements. 

\subsubsection{Supplementary guidance} In different use cases, question analysis may reveal categories that are not presented in XAI Question Bank. It is worth noting that XAI Question Bank was developed with use cases mainly in the area of AI decision-support for business users.\cite{liao2020questioning}. So new question categories are especially likely to appear for AI systems that are out of the scope of performing a standard decision-support task by providing predictions or recommendations. For example, in the crime analytics use case where we tested the design process, users were not only receiving AI predictions but also configuring their own model. New categories of questions such as regarding \textit{data requirements} for the model to handle, and the \textit{configuration process} appeared. Understanding the intended scope of the AI system early on could help preparing for the Question Analysis step.

For the prioritization task, a group could also vote on the question categories to focus based on both the insights gained from the user research and other criteria the team share. A more formal approach to actively engage the team to form a shared vision is to vote on a prioritization matrix, where one axis is about user value and the other is about feasibility, and the goal is to identify question categories that are both important for the UX and relatively easy to deliver. To complete this group voting, however, may require the team to have a level of XAI expertise to be able to effectively assess the feasibility.

\subsection{Step 3: Mapping Questions to Modeling Solutions}
Step 3 is for designers and AI engineers to take the prioritized question categories and start identifying an initial set of XAI techniques as functional elements that the design should include. While how a team choose to accomplish the task is flexible, we provide a mapping guide in Figure~\ref{fig:mapping} to support the designer-engineer collaboration. For each question category, the guide suggests examples of explanations to answer the question, grounded in available XAI techniques, to help designers form an understanding of what should be done and what is possible to explain (\textbf{R2}). We also link to popular code libraries for suggested XAI techniques, so that AI engineers not only have a set of concrete techniques to consider but also access to technical details. This kind of mapping artifact could serve as a boundary object to ground the discussions and consensus building between AI engineers and designers (\textbf{R3}).

\subsubsection{Activities in the use case}
A group of designers and AI engineers held multiple meetings to discuss the findings from the first two steps and identify candidate XAI technical solutions. The four priority question categories were used to guide the search, with an early version of the mapping guide in Figure~\ref{fig:mapping}. The identified user requirements also provided guidance for several decision points. Grounded in the user questions and the suggested mapping, on one side,  AI engineers were able to consider the characteristics of the underlying model, on the other side, designers were able to provide input based on a deeper understanding about the users and usage contexts, both of which helped narrowing down the choices of XAI solutions.

\subsubsection{Outcomes in the use case} Figure~\ref{fig:interface} is a version of the design we created in Step 4, with XAI features marked with corresponding question categories. Step 3 focused on discussions to identify functional elements without yet creating the actual design. However, for the easiness of reading, we will refer to Figure~\ref{fig:interface} for illustration purpose when we discuss specific XAI techniques.

For \textit{Why} questions, based on Figure~\ref{fig:mapping}, the team quickly converged on the idea of explaining by ``describing what key features of the particular instance determine the model's prediction on it'' (output would look like part 3 in Figure~\ref{fig:interface}) because participants were explicitly interested in patients' ``risk factors''. While Figure~\ref{fig:mapping} suggested code libraries for state-of-the-art post-hoc XAI techniques including LIME~\cite{ribeiro2016should} and SHAP~\cite{lundberg2017unified}, the team opted for a Feature Importance Analysis approach that the team could re-use code from a previous project. 

For \textit{How to be that} questions, the team chose to use an algorithm called Contrastive Explanation Method (CEM)~\cite{dhurandhar2018explanations}. As suggested in the mapping guide, CEM identifies features that if changed (increased, decreased, absent or present) could lead to the highest changes in the prediction. CEM not only identifies the top features to change, but also quantifies each changed feature's impact. While the top features to change could help satisfy \textit{UR2: Help me determine the most actionable next steps for the management of patients}, the quantification could potentially guide users' actions to satisfy \textit{UR3: Help me decide whether I should spend more time on further evaluation or management of a patient}--if the impact is small, it may not be worth engaging with interventions. The direct output would look like part 5 in Figure~\ref{fig:interface}.

For \textit{Data} questions, the answers are usually considered factual explanations about the training data of the model. The team carefully examined what were being asked in elicited questions and consulted the mapping guide, and decided to present facts about the data source, size of the data, distributions of key demographics, and the limitations of the data, specifically what common datasets for this kind of system are not included, as illustrated in the pop-up panel in part 1 of Figure~\ref{fig:interface}.

For \textit{Performance} questions, as discussed we noticed that the elicit questions were mostly regarding the limitations of the AI, i.e. in what situations its predictions are less reliable. The AI engineers recognized while it is possible to generate high-level descriptions about these limitations, their generalizability and utility might be questionable. The team converged on the idea of presenting the model's uncertainty level together with each prediction, so users could be alerted for cases that they should not rely on the AI's prediction.

In summary, the team reached a consensus to present in the interface feature importance information (\textit{Why}), Contrastive explanation as generated by CEM (\textit{How to be that}), multiple facts about the training data and confidence information. At this point, designers were ready to create an initial design including those elements, and AI engineers could start exploring the implementation in Step 4.

\subsubsection{Supplementary guidance} We acknowledge that the team had a comparatively high expertise on XAI, with the involvement of several XAI researchers. It is possible in teams where the AI engineers were less informed on the topic the collaborative problem-solving and the understanding on the mapping guide would be more challenging. A team could benefit from additional training or educational materials to understand the technical landscape of XAI, and the strengths and limitations of different XAI algorithms. In the slideshow (see supplementary material), we provide an overview of different XAI techniques and provide links to tutorials on the topic. 

In this project, an RNN model was pre-selected for its performance advantage. That limited our choices of XAI techniques, for example, only post-hoc techniques like CEM can be used. For projects in an early stage, user needs for explainability could impact not only the choice of XAI technique but also the model itself, such as choosing a type of model that can generate a desired form of explanation even at the cost of performance (e.g., using a simpler, directly interpretable model).

\subsection{Step 4: Iterative Design and Evaluation}
~\label{step4}
In the last step, the XAI UX design is iteratively developed and evaluated with the user requirements identified in Step 2. It is important for designers and AI engineers to continue collaborating with frequent touching points (\textbf{R3}), because the translation of XAI techniques into XAI UX would often require in-parallel refinement of models or XAI techniques (\textbf{R4}). During the design iterations, it is also preferable to have target users involved to provide feedback.

This step starts with designers creating an initial version of the design that include the XAI elements identified in Step 3. Designers can refer to exemplars in the mapping guide and other references (e.g., illustrations in the linked XAI toolkits) for the chosen XAI techniques to explore the design space. Data scientists can provide additional examplers of algorithmic output to support the exploration of design.

Then the design is evaluated by enumerating on the user requirements identified in Step 2. In an early design stage, designers and the team can choose to engage in heuristic evaluation to identify gaps. In a later stage, it would be beneficial to conduct evaluative user research. One approach we found helpful is to ask participants of user research, after they experience the design prototype, to explicitly rate on each user requirements and reflect on the gaps. The gaps will then guide the next iteration of design and technical development.

It is also possible to use identified user requirements to define success metrics that guide the development of models and XAI techniques. For example, if efficiency or low cognitive workload is a primary user requirement, the team should prioritize improving the output compactness of the XAI technique.

\subsubsection{Activities in the use case}
\begin{figure*}
  \centering
  \includegraphics[width=1\columnwidth]{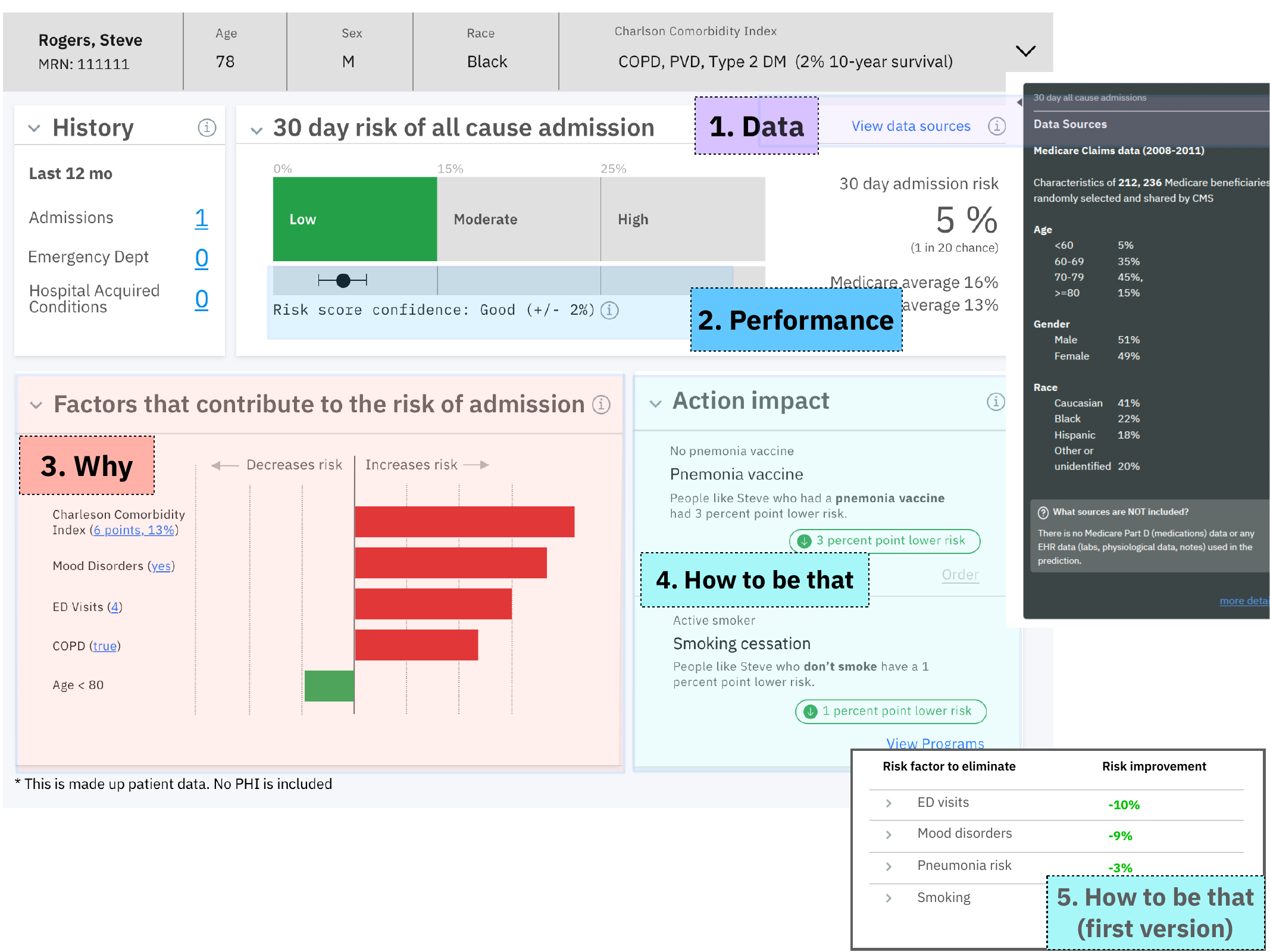}
   \vspace{-2em}
  \caption{Design prototype created by following the design process, with explainability features marked with corresponding user questions they answer. Part 1: \textit{Data} explanations with a hover-over pop-up window. Part 2: \textit{Performance} explanation with confidence information. Part 3: \textit{Why} explanation with feature importance visualization. Part 4: \textit{How to be that} explanation with ``risk factors to eliminate'' generated by Contrastive Explanation Method (CEM) then linked to actionable resources. Part 5: original \textit{How to be that} explanation, with direct output of CEM, but lacking actionability  }~\label{fig:interface}

\end{figure*}
After the initial design was created, it first went through iterations within the team, by reflecting on the user requirements as heuristic evaluation and discussing gaps to fill. Designers conducted evaluative research with the same group of participants and received feedback. Designers then shared the insights with AI engineers and the rest of the team to guide further improvement of the AI system. In this paper we will leave out the final system and last iteration of user research for proprietary reasons, but present an example in the first round of iteration that the team engaged in internally to illustrate the design process.

\subsubsection{Outcomes in the use case}
Figure~\ref{fig:interface} presents a design prototype of the interface for the AI system predicting healthcare adverse events, presented with fictional patient data. Different areas in Figure~\ref{fig:interface} are marked with user question types they correspond to. 

One place to illustrate the outcome of an iterative design process is the contrast between part 4 and part 5, the latter of which is the first version of design to present contrastive explanations (output of CEM algorithm). It lists the top features that if changed could lower the patient's predicted risk (\textit{How to be that}), as well as the quantity of the impact of the change on the predicted risk. However, when evaluated against \textit{UR2: Help me determine the most actionable next steps for the management of patients}, this design was found to be critically lacking actionablity. First, the output of CEM algorithm, when suggesting top features to change (i.e. risk factors to eliminate), does not consider the practical changeability of the features or cost of changes, such as demographic factors or one's medical history (e.g., ED visit and Mood disorders as in part 5 of Figure~\ref{fig:interface}). Second, the suggested risk factors only indicate ``what to change'', but not ``how to change''.   Designers could recall a quote from a participant (P4) ``\textit{I would be looking for AI's suggestion...is it plausible? Is it actionable? It needs to be associated with reasonable actions, and also with reasonable cost. I can't be like, calling everyone born in July}''

The team reflected on these limitations and came up with a new version of design (part 4 in Figure~\ref{fig:interface}). The improvement first required AI engineers to update the CEM algorithm, putting constraints to prevent suggesting of a group of features that are considered unchangeable. Second, data scientists and designers worked together to identify a list of frequently appearing features in CEM output. With the input of experts in the healthcare domain, the team then created an additional layer of output by linking these features to external guidelines or resources for ``how to change'', whenever possible. For example, in part 4 of Figure~\ref{fig:interface}, the suggested risk factor to eliminate, ``pneumonia risk'' was linked to information about pneumonia vaccine and link to order. ``Smoking'' was linked to information about smoking cessation and links to supportive programs. 

For the rest of the design prototype, it answers the \textit{Why} question with a visualization of feature importance (Part 3), which ranks features of the patient by their contribution to the patient's predicted risk, and visualizes the quantity and the direction (increase/decrease the risk) of the impact. It answers the \textit{Data} question with a pop-up window when hovering over that presents facts about the training data, as identified in Step 3 (Part 1). Lastly, the confidence information was presented together with the the risk prediction, including both a verbal description and an error bar visualization.

\subsubsection{Supplementary guidance} Most points were discussed in the beginning of Section~\ref{step4}. We call attention to the possibilities that new user questions, user requirements and the needs to add additional XAI technique can appear during evaluation, and should be incorporated into future iterations.

\section{Discussions}
We first discuss some practical considerations and feedback on the adoption of this design process, then reflect on lesson learned for tackling unique challenges in designing AI UX.
 
\subsection{Considerations and feedback for adoption}
Since we started developing this work, we were confronted with the challenge of how to motivate and enable product teams with varying degrees of knowledge on the topic of XAI to adopt the design process. We decided to host workshops to bring together different roles in a product team, including designers and AI engineers, to introduce the design process in an interactive environment. A workshop would also allow us to provide richer materials for the product team to first form a shared understanding on the space of explainable AI and the possible outcomes of XAI UX. The supplementary material of this paper includes a slideshow we are using in these workshops. Following the introductory material, a team could choose to work on their own use case at the workshop. For example, a team could invite a small number of users, complete the question elicitation, work together on the question analysis and mapping to modeling solutions, then leave the workshop with a set of identified functional elements to move forward in design (Step 3). This approach would be especially useful for product teams that are interested in accelerating their development of explainability features. A team could also learn about the process through the use case we provide, then make plans to carry out the design process for their own product. Alternatively, the slideshow can be used as a reading material and followed as step-by-step instructions.

We received positive feedback from the designers and AI engineers we involved in the method development process. First, they appreciated the opportunity to \textit{introduce a product team to the topic of XAI in a holistic way} through an overview of the technical and design spaces, concrete examples of possible design outcomes, and an actionable method to follow. One data scientist commented that ``\textit{XAI is one of those hyped topics but [what] everyone talks about is LIME and SHAP}''. This comment highlights the potential ``availability bias'' in practitioners' choice of XAI techniques without due attention to their suitability for users. Introduction to an XAI specific design process could help product teams understand the technical space of XAI holistically and examine the choices thoroughly. The proposed method is scalable to the addition of new XAI techniques, as they could be added to the mapping guide, while the elicitation and understanding of user questions are universally applicable. 

Second, designers we consulted with appreciated the \textit{core notion of putting users' questions first}. One designer commented that even if some product teams cannot execute the design process as instructed, or they may choose to adapt it, ``being able to recognize that we need to know user questions and address them goes a long way''. This comment summarizes our intention well. Instead of suggesting complex guidelines on how to design different types of XAI systems, we propose to ground the solutions to when, where and how to provide AI explanations empirically in the types of user questions asked. By putting user questions in the forefront, this process centers the design and the team's shared vision for XAI UX around users instead of the AI algorithm or modeling solutions. 

Lastly, many recognized the benefits of \textit{promoting closer collaboration between designers and AI engineers}, as ``a rare opportunity to bring data scientists further into the design process''. Such responses resonate with prior work~\cite{yang2018investigating,yang2019profiling,girardin2017user,liao2020questioning,dove2017ux} highlighting the unique needs for AI UX design to have in-parallel development of design and modeling solutions, and challenges for designers and data scientists to collaborate due to communication barriers and a lack of shared workflows. In particular, our proposed method tackles the pain points of designer-engineer collaboration with two measures: a structured workflow with concrete touch points, and boundary objects (the question mapping guide) to bridge user needs and technical solutions. Moving forward, we hope this XAI design process provides a ``petri dish'' to explore how to better support designer-engineer co-creation for AI UX.   

\subsection{User-centered AI design: guiding the use of toolboxes of design materials}
As discussed earlier, explainability is one of many AI specific design issues that are enabled and bounded by a growing space of technical solutions. In a similar vein, design issues such as fairness and uncertainty may also face the challenges of selection and translation to use a toolbox of algorithmic solutions. For example, there are a growing collection of techniques to quantify model fairness, identify areas of biases and provide de-biasing interventions~\cite{bellamy2018ai}. The challenge of selection, to identify appropriate fairness metrics and intervention methods to use, should rely on a careful understanding on the fairness notion in a specific domain, and how the choice may impact stakeholders and the community at large. Furthermore, how to communicate output of fairness metrics to users who may be AI novice, have low numeracy, or deviate from normative views of fairness, remains a challenge of translation. For model uncertainty, both the properties of the underlying model and user characteristics should guide the selection of both the algorithm to quantify uncertainty and the method to communicate uncertainty~\cite{bhatt2020uncertainty}. 

More broadly, in current AI product practices, the choices of model and various decisions in the model development process are often dominated by technical considerations such as model performance metrics. This may pose problems down the road if the chosen model lacks means to deliver desired elements in user experiences. We believe there is a need to innovate the design processes of AI products by formulating shared design vision early on to guide the choices of modeling solutions as well as the success metrics. To advocate for "design first" practices calls for a change of perspective by viewing the collection of AI algorithms as ``a toolbox of design materials''.

HCI and design researchers have discussed extensively challenges working with AI as a new and difficult design material~\cite{dove2017ux,yang2018investigating}, including a lack of clear understanding of AI's capabilities, difficulties in envisioning uses that do not yet exist, and obstacles in prototyping. We further add to the challenges designing AI UX with a toolbox of design materials, where a collection of AI techniques are available but have different strengths, weakness and appropriate usage contexts. To form a designerly understanding of a toolbox requires not only understanding the capabilities and limitations of individual tools, but also effective ways to navigate the technical space. To tackle these challenges requires developing more actionable frameworks to organize the toolbox. For instance, there exist many taxonomies to classify XAI algorithms~\cite{arrieta2020explainable,guidotti2018survey} based on the technical details such as the stage of model development or type of model and data. However, they present significant disconnect from both the design goals framed as values to the users, and what is required to have a designerly understanding of the toolbox. 

Our work re-frames and organizes the toolbox of XAI techniques according to the type of user questions they can answer, which could be directly linked to output of user research. This re-framing would encourage product teams, in order to navigate the toolbox of XAI algorithms, to foreground user needs and requirements instead of technical promises. More fundamentally, we believe a shift needs to happen to re-frame technical spaces of AI technologies by the human needs, values, and conditions they serve. This shift of views should start from the roots, by carefully contextualizing AI algorithms developed in academic work, articulating their strengths and weakness, and examining their potential impact for different types of users and usage contexts. 


\section{Limitations}
The Question-Driven Design Process for XAI was developed based on feedback from 21 AI practitioners and tested with two use cases. However, we acknowledge the limitation that these practitioners worked for the same large technology company, although they were from various product lines and locations. It is possible that design practices are different in other organizations. It is also possible that this design process would not be able to account for nuances in certain AI products and can break down in certain design environments. We invite researchers and practitioners to experiment with and further refine the design process. 

We are not claiming this should be the only approach to XAI UX design. Other methods have been proposed in the literature~\cite{eiband2018bringing,wolf2019explainability}, which offer alternative means to address the \textit{what to explain} and \textit{how to explain} questions. Our approach also does not offer normative guidelines on how XAI UX should be designed or evaluated. It is up to a product team to explore the design solutions and determine when the design is good enough without requiring further iterations. Future work could explore developing standard evaluation methods and metrics for XAI UX, to complement the set of user requirements identified from user research. 

Lastly, as discussed, the mapping guide we provided is based on the XAI Question Bank~\cite{liao2020questioning}. In practices, especially for AI systems outside that scope of a standard decision-support ML system, user questions can fall outside the question bank and the mapping guide, which requires additional effort to seek solutions by collaborating with AI engineers. The guide also does not account for technical nuances. For example, there could be reliability, safety and ethical concerns to use post-hoc XAI techniques~\cite{rudin2019stop} in certain contexts. These are important discussions that a team should have. In short, the XAI Question Bank and mapping guide should be used as a suggestion only and a work-in-progress repository that should be continuously expanded and refined. 

\section{Conclusion}
One of the critical and pervasive design issues of AI systems is their explainability--how to provide appropriate information to help users understand the AI's functions and decisions. Meanwhile, the technical field of explainable AI (XAI) has experienced a remarkable surge in recent years and produced a rich toolbox of XAI techniques. To facilitate the utilization of this toolbox, especially to \textit{select} suitable XAI techniques and \textit{translate} the algorithmic output into UX solutions that effectively serve user needs in real-world AI systems, we propose a Question-Driven Design Process for XAI. The method was built upon prior literature on supporting design of AI and XAI systems~\cite{yang2018investigating,yang2019profiling,girardin2017user,dove2017ux,liao2020questioning,eiband2018bringing}, based on feedback from 21 AI practitioners, and tested and refined with two use cases. Specifically, the design process centers XAI UX design around users by grounding the user needs and requirements, design, and evaluation criteria in users' questions to understand the AI. The process supports a designerly understanding on the affordance of different XAI techniques by providing a mapping guide between prototypical user questions and exemplars of XAI techniques to answer these questions. The design process also aims to support designer-AI-engineer collaboration, as well as parallel refinement of design and technical solutions, through a shared workflow and using the mapping guide as boundary objects for consensus building. Our work offers both an actionable framework for XAI UX design, and a test-bed to address challenges in designing AI systems with a collection of algorithmic solutions as a toolbox of design materials.

\bibliographystyle{ACM-Reference-Format}
\bibliography{sample-base}


\begin{thebibliography}{60}


\ifx \showCODEN    \undefined \def \showCODEN     #1{\unskip}     \fi
\ifx \showDOI      \undefined \def \showDOI       #1{#1}\fi
\ifx \showISBNx    \undefined \def \showISBNx     #1{\unskip}     \fi
\ifx \showISBNxiii \undefined \def \showISBNxiii  #1{\unskip}     \fi
\ifx \showISSN     \undefined \def \showISSN      #1{\unskip}     \fi
\ifx \showLCCN     \undefined \def \showLCCN      #1{\unskip}     \fi
\ifx \shownote     \undefined \def \shownote      #1{#1}          \fi
\ifx \showarticletitle \undefined \def \showarticletitle #1{#1}   \fi
\ifx \showURL      \undefined \def \showURL       {\relax}        \fi
\providecommand\bibfield[2]{#2}
\providecommand\bibinfo[2]{#2}
\providecommand\natexlab[1]{#1}
\providecommand\showeprint[2][]{arXiv:#2}

\bibitem[\protect\citeauthoryear{??}{H2o}{2017}]%
        {H2o}
 \bibinfo{year}{2017}\natexlab{}.
\newblock \bibinfo{title}{H2O.ai Machine Learning Interpretability}.
\newblock
\newblock
\newblock
\shownote{https://github.com/h2oai/mli-resources.}


\bibitem[\protect\citeauthoryear{??}{mod}{2019}]%
        {model_card}
 \bibinfo{year}{2019}\natexlab{}.
\newblock \bibinfo{title}{Google Cloud Model Cards}.
\newblock
\newblock
\newblock
\shownote{https://modelcards.withgoogle.com/about.}


\bibitem[\protect\citeauthoryear{??}{AIX}{2019}]%
        {AIX}
 \bibinfo{year}{2019}\natexlab{}.
\newblock \bibinfo{title}{IBM AIX 360}.
\newblock
\newblock
\newblock
\shownote{aix360.mybluemix.net/.}


\bibitem[\protect\citeauthoryear{??}{Mic}{2019}]%
        {Microsoft}
 \bibinfo{year}{2019}\natexlab{}.
\newblock \bibinfo{title}{Microsoft InterpretML}.
\newblock
\newblock
\newblock
\shownote{hhttps://github.com/interpretml/interpret.}


\bibitem[\protect\citeauthoryear{??}{ali}{2019}]%
        {alibi}
 \bibinfo{year}{2019}\natexlab{}.
\newblock \bibinfo{title}{SeldonIO Alibi Explain}.
\newblock
\newblock
\newblock
\shownote{https://docs.seldon.io/projects/alibi/en/stable/.}


\bibitem[\protect\citeauthoryear{??}{fac}{2020}]%
        {factsheets}
 \bibinfo{year}{2020}\natexlab{}.
\newblock \bibinfo{title}{IBM AI FactSheets 360}.
\newblock
\newblock
\newblock
\shownote{https://aifs360.mybluemix.net/.}


\bibitem[\protect\citeauthoryear{Abdul, Vermeulen, Wang, Lim, and
  Kankanhalli}{Abdul et~al\mbox{.}}{2018}]%
        {abdul2018trends}
\bibfield{author}{\bibinfo{person}{Ashraf Abdul}, \bibinfo{person}{Jo
  Vermeulen}, \bibinfo{person}{Danding Wang}, \bibinfo{person}{Brian~Y Lim},
  {and} \bibinfo{person}{Mohan Kankanhalli}.} \bibinfo{year}{2018}\natexlab{}.
\newblock \showarticletitle{Trends and trajectories for explainable,
  accountable and intelligible systems: An hci research agenda}. In
  \bibinfo{booktitle}{\emph{Proceedings of the 2018 CHI conference on human
  factors in computing systems}}. \bibinfo{pages}{1--18}.
\newblock


\bibitem[\protect\citeauthoryear{Adadi and Berrada}{Adadi and Berrada}{2018}]%
        {adadi2018peeking}
\bibfield{author}{\bibinfo{person}{Amina Adadi} {and} \bibinfo{person}{Mohammed
  Berrada}.} \bibinfo{year}{2018}\natexlab{}.
\newblock \showarticletitle{Peeking inside the black-box: a survey on
  explainable artificial intelligence (XAI)}.
\newblock \bibinfo{journal}{\emph{IEEE access}}  \bibinfo{volume}{6}
  (\bibinfo{year}{2018}), \bibinfo{pages}{52138--52160}.
\newblock


\bibitem[\protect\citeauthoryear{Alqaraawi, Schuessler, Wei{\ss}, Costanza, and
  Berthouze}{Alqaraawi et~al\mbox{.}}{2020}]%
        {alqaraawi2020evaluating}
\bibfield{author}{\bibinfo{person}{Ahmed Alqaraawi}, \bibinfo{person}{Martin
  Schuessler}, \bibinfo{person}{Philipp Wei{\ss}}, \bibinfo{person}{Enrico
  Costanza}, {and} \bibinfo{person}{Nadia Berthouze}.}
  \bibinfo{year}{2020}\natexlab{}.
\newblock \showarticletitle{Evaluating saliency map explanations for
  convolutional neural networks: a user study}. In
  \bibinfo{booktitle}{\emph{Proceedings of the 25th International Conference on
  Intelligent User Interfaces}}. \bibinfo{pages}{275--285}.
\newblock


\bibitem[\protect\citeauthoryear{Amershi, Weld, Vorvoreanu, Fourney, Nushi,
  Collisson, Suh, Iqbal, Bennett, Inkpen, et~al\mbox{.}}{Amershi
  et~al\mbox{.}}{2019}]%
        {amershi2019guidelines}
\bibfield{author}{\bibinfo{person}{Saleema Amershi}, \bibinfo{person}{Dan
  Weld}, \bibinfo{person}{Mihaela Vorvoreanu}, \bibinfo{person}{Adam Fourney},
  \bibinfo{person}{Besmira Nushi}, \bibinfo{person}{Penny Collisson},
  \bibinfo{person}{Jina Suh}, \bibinfo{person}{Shamsi Iqbal},
  \bibinfo{person}{Paul~N Bennett}, \bibinfo{person}{Kori Inkpen},
  {et~al\mbox{.}}} \bibinfo{year}{2019}\natexlab{}.
\newblock \showarticletitle{Guidelines for human-AI interaction}. In
  \bibinfo{booktitle}{\emph{Proceedings of the 2019 chi conference on human
  factors in computing systems}}. \bibinfo{pages}{1--13}.
\newblock


\bibitem[\protect\citeauthoryear{Arrieta, D{\'\i}az-Rodr{\'\i}guez, Del~Ser,
  Bennetot, Tabik, Barbado, Garc{\'\i}a, Gil-L{\'o}pez, Molina, Benjamins,
  et~al\mbox{.}}{Arrieta et~al\mbox{.}}{2020}]%
        {arrieta2020explainable}
\bibfield{author}{\bibinfo{person}{Alejandro~Barredo Arrieta},
  \bibinfo{person}{Natalia D{\'\i}az-Rodr{\'\i}guez}, \bibinfo{person}{Javier
  Del~Ser}, \bibinfo{person}{Adrien Bennetot}, \bibinfo{person}{Siham Tabik},
  \bibinfo{person}{Alberto Barbado}, \bibinfo{person}{Salvador Garc{\'\i}a},
  \bibinfo{person}{Sergio Gil-L{\'o}pez}, \bibinfo{person}{Daniel Molina},
  \bibinfo{person}{Richard Benjamins}, {et~al\mbox{.}}}
  \bibinfo{year}{2020}\natexlab{}.
\newblock \showarticletitle{Explainable Artificial Intelligence (XAI):
  Concepts, taxonomies, opportunities and challenges toward responsible AI}.
\newblock \bibinfo{journal}{\emph{Information Fusion}}  \bibinfo{volume}{58}
  (\bibinfo{year}{2020}), \bibinfo{pages}{82--115}.
\newblock


\bibitem[\protect\citeauthoryear{Arya, Bellamy, Chen, Dhurandhar, Hind,
  Hoffman, Houde, Liao, Luss, Mojsilovi{\'c}, et~al\mbox{.}}{Arya
  et~al\mbox{.}}{2019}]%
        {arya2019one}
\bibfield{author}{\bibinfo{person}{Vijay Arya}, \bibinfo{person}{Rachel~KE
  Bellamy}, \bibinfo{person}{Pin-Yu Chen}, \bibinfo{person}{Amit Dhurandhar},
  \bibinfo{person}{Michael Hind}, \bibinfo{person}{Samuel~C Hoffman},
  \bibinfo{person}{Stephanie Houde}, \bibinfo{person}{Q~Vera Liao},
  \bibinfo{person}{Ronny Luss}, \bibinfo{person}{Aleksandra Mojsilovi{\'c}},
  {et~al\mbox{.}}} \bibinfo{year}{2019}\natexlab{}.
\newblock \showarticletitle{One explanation does not fit all: A toolkit and
  taxonomy of ai explainability techniques}.
\newblock \bibinfo{journal}{\emph{arXiv preprint arXiv:1909.03012}}
  (\bibinfo{year}{2019}).
\newblock


\bibitem[\protect\citeauthoryear{Begel, Tang, Andrist, Barnett, Carbary,
  Choudhury, Cutrell, Fung, Junuzovic, McDuff, et~al\mbox{.}}{Begel
  et~al\mbox{.}}{2020}]%
        {begel2020lessons}
\bibfield{author}{\bibinfo{person}{Andrew Begel}, \bibinfo{person}{John Tang},
  \bibinfo{person}{Sean Andrist}, \bibinfo{person}{Michael Barnett},
  \bibinfo{person}{Tony Carbary}, \bibinfo{person}{Piali Choudhury},
  \bibinfo{person}{Edward Cutrell}, \bibinfo{person}{Alberto Fung},
  \bibinfo{person}{Sasa Junuzovic}, \bibinfo{person}{Daniel McDuff},
  {et~al\mbox{.}}} \bibinfo{year}{2020}\natexlab{}.
\newblock \showarticletitle{Lessons Learned in Designing AI for Autistic
  Adults}. In \bibinfo{booktitle}{\emph{The 22nd International ACM SIGACCESS
  Conference on Computers and Accessibility}}. \bibinfo{pages}{1--6}.
\newblock


\bibitem[\protect\citeauthoryear{Bellamy, Dey, Hind, Hoffman, Houde, Kannan,
  Lohia, Martino, Mehta, Mojsilovic, et~al\mbox{.}}{Bellamy
  et~al\mbox{.}}{2018}]%
        {bellamy2018ai}
\bibfield{author}{\bibinfo{person}{Rachel~KE Bellamy}, \bibinfo{person}{Kuntal
  Dey}, \bibinfo{person}{Michael Hind}, \bibinfo{person}{Samuel~C Hoffman},
  \bibinfo{person}{Stephanie Houde}, \bibinfo{person}{Kalapriya Kannan},
  \bibinfo{person}{Pranay Lohia}, \bibinfo{person}{Jacquelyn Martino},
  \bibinfo{person}{Sameep Mehta}, \bibinfo{person}{Aleksandra Mojsilovic},
  {et~al\mbox{.}}} \bibinfo{year}{2018}\natexlab{}.
\newblock \showarticletitle{AI Fairness 360: An extensible toolkit for
  detecting, understanding, and mitigating unwanted algorithmic bias}.
\newblock \bibinfo{journal}{\emph{arXiv preprint arXiv:1810.01943}}
  (\bibinfo{year}{2018}).
\newblock


\bibitem[\protect\citeauthoryear{Bhatt, Xiang, Sharma, Weller, Taly, Jia,
  Ghosh, Puri, Moura, and Eckersley}{Bhatt et~al\mbox{.}}{2020a}]%
        {bhatt2020explainable}
\bibfield{author}{\bibinfo{person}{Umang Bhatt}, \bibinfo{person}{Alice Xiang},
  \bibinfo{person}{Shubham Sharma}, \bibinfo{person}{Adrian Weller},
  \bibinfo{person}{Ankur Taly}, \bibinfo{person}{Yunhan Jia},
  \bibinfo{person}{Joydeep Ghosh}, \bibinfo{person}{Ruchir Puri},
  \bibinfo{person}{Jos{\'e}~MF Moura}, {and} \bibinfo{person}{Peter
  Eckersley}.} \bibinfo{year}{2020}\natexlab{a}.
\newblock \showarticletitle{Explainable machine learning in deployment}. In
  \bibinfo{booktitle}{\emph{Proceedings of the 2020 Conference on Fairness,
  Accountability, and Transparency}}. \bibinfo{pages}{648--657}.
\newblock


\bibitem[\protect\citeauthoryear{Bhatt, Zhang, Antor{\'a}n, Liao, Sattigeri,
  Fogliato, Melan{\c{c}}on, Krishnan, Stanley, Tickoo, et~al\mbox{.}}{Bhatt
  et~al\mbox{.}}{2020b}]%
        {bhatt2020uncertainty}
\bibfield{author}{\bibinfo{person}{Umang Bhatt}, \bibinfo{person}{Yunfeng
  Zhang}, \bibinfo{person}{Javier Antor{\'a}n}, \bibinfo{person}{Q~Vera Liao},
  \bibinfo{person}{Prasanna Sattigeri}, \bibinfo{person}{Riccardo Fogliato},
  \bibinfo{person}{Gabrielle~Gauthier Melan{\c{c}}on},
  \bibinfo{person}{Ranganath Krishnan}, \bibinfo{person}{Jason Stanley},
  \bibinfo{person}{Omesh Tickoo}, {et~al\mbox{.}}}
  \bibinfo{year}{2020}\natexlab{b}.
\newblock \showarticletitle{Uncertainty as a Form of Transparency: Measuring,
  Communicating, and Using Uncertainty}.
\newblock \bibinfo{journal}{\emph{arXiv preprint arXiv:2011.07586}}
  (\bibinfo{year}{2020}).
\newblock


\bibitem[\protect\citeauthoryear{Bogers, Frens, Van~Kollenburg, Deckers, and
  Hummels}{Bogers et~al\mbox{.}}{2016}]%
        {bogers2016connected}
\bibfield{author}{\bibinfo{person}{Sander Bogers}, \bibinfo{person}{Joep
  Frens}, \bibinfo{person}{Janne Van~Kollenburg}, \bibinfo{person}{Eva
  Deckers}, {and} \bibinfo{person}{Caroline Hummels}.}
  \bibinfo{year}{2016}\natexlab{}.
\newblock \showarticletitle{Connected baby bottle: A design case study towards
  a framework for data-enabled design}. In
  \bibinfo{booktitle}{\emph{Proceedings of the 2016 ACM Conference on Designing
  Interactive Systems}}. \bibinfo{pages}{301--311}.
\newblock


\bibitem[\protect\citeauthoryear{Bu{\c{c}}inca, Lin, Gajos, and
  Glassman}{Bu{\c{c}}inca et~al\mbox{.}}{2020}]%
        {buccinca2020proxy}
\bibfield{author}{\bibinfo{person}{Zana Bu{\c{c}}inca}, \bibinfo{person}{Phoebe
  Lin}, \bibinfo{person}{Krzysztof~Z Gajos}, {and} \bibinfo{person}{Elena~L
  Glassman}.} \bibinfo{year}{2020}\natexlab{}.
\newblock \showarticletitle{Proxy tasks and subjective measures can be
  misleading in evaluating explainable ai systems}. In
  \bibinfo{booktitle}{\emph{Proceedings of the 25th International Conference on
  Intelligent User Interfaces}}. \bibinfo{pages}{454--464}.
\newblock


\bibitem[\protect\citeauthoryear{Cai, Jongejan, and Holbrook}{Cai
  et~al\mbox{.}}{2019}]%
        {cai2019effects}
\bibfield{author}{\bibinfo{person}{Carrie~J Cai}, \bibinfo{person}{Jonas
  Jongejan}, {and} \bibinfo{person}{Jess Holbrook}.}
  \bibinfo{year}{2019}\natexlab{}.
\newblock \showarticletitle{The effects of example-based explanations in a
  machine learning interface}. In \bibinfo{booktitle}{\emph{Proceedings of the
  24th International Conference on Intelligent User Interfaces}}.
  \bibinfo{pages}{258--262}.
\newblock


\bibitem[\protect\citeauthoryear{Chakraborti, Sreedharan, and
  Kambhampati}{Chakraborti et~al\mbox{.}}{2020}]%
        {chakraborti2020emerging}
\bibfield{author}{\bibinfo{person}{Tathagata Chakraborti},
  \bibinfo{person}{Sarath Sreedharan}, {and} \bibinfo{person}{Subbarao
  Kambhampati}.} \bibinfo{year}{2020}\natexlab{}.
\newblock \showarticletitle{The emerging landscape of explainable ai planning
  and decision making}.
\newblock \bibinfo{journal}{\emph{arXiv preprint arXiv:2002.11697}}
  (\bibinfo{year}{2020}).
\newblock


\bibitem[\protect\citeauthoryear{Cheng, Wang, Zhang, O'Connell, Gray, Harper,
  and Zhu}{Cheng et~al\mbox{.}}{2019}]%
        {cheng2019explaining}
\bibfield{author}{\bibinfo{person}{Hao-Fei Cheng}, \bibinfo{person}{Ruotong
  Wang}, \bibinfo{person}{Zheng Zhang}, \bibinfo{person}{Fiona O'Connell},
  \bibinfo{person}{Terrance Gray}, \bibinfo{person}{F~Maxwell Harper}, {and}
  \bibinfo{person}{Haiyi Zhu}.} \bibinfo{year}{2019}\natexlab{}.
\newblock \showarticletitle{Explaining decision-making algorithms through UI:
  Strategies to help non-expert stakeholders}. In
  \bibinfo{booktitle}{\emph{Proceedings of the 2019 chi conference on human
  factors in computing systems}}. \bibinfo{pages}{1--12}.
\newblock


\bibitem[\protect\citeauthoryear{Dhurandhar, Chen, Luss, Tu, Ting, Shanmugam,
  and Das}{Dhurandhar et~al\mbox{.}}{2018}]%
        {dhurandhar2018explanations}
\bibfield{author}{\bibinfo{person}{Amit Dhurandhar}, \bibinfo{person}{Pin-Yu
  Chen}, \bibinfo{person}{Ronny Luss}, \bibinfo{person}{Chun-Chen Tu},
  \bibinfo{person}{Paishun Ting}, \bibinfo{person}{Karthikeyan Shanmugam},
  {and} \bibinfo{person}{Payel Das}.} \bibinfo{year}{2018}\natexlab{}.
\newblock \showarticletitle{Explanations based on the missing: towards
  contrastive explanations with pertinent negatives}. In
  \bibinfo{booktitle}{\emph{Proceedings of the 32nd International Conference on
  Neural Information Processing Systems}}. \bibinfo{pages}{590--601}.
\newblock


\bibitem[\protect\citeauthoryear{Dodge, Liao, Zhang, Bellamy, and Dugan}{Dodge
  et~al\mbox{.}}{2019}]%
        {dodge2019explaining}
\bibfield{author}{\bibinfo{person}{Jonathan Dodge}, \bibinfo{person}{Q~Vera
  Liao}, \bibinfo{person}{Yunfeng Zhang}, \bibinfo{person}{Rachel~KE Bellamy},
  {and} \bibinfo{person}{Casey Dugan}.} \bibinfo{year}{2019}\natexlab{}.
\newblock \showarticletitle{Explaining models: an empirical study of how
  explanations impact fairness judgment}. In
  \bibinfo{booktitle}{\emph{Proceedings of the 24th International Conference on
  Intelligent User Interfaces}}. \bibinfo{pages}{275--285}.
\newblock


\bibitem[\protect\citeauthoryear{Doshi-Velez and Kim}{Doshi-Velez and
  Kim}{2017}]%
        {doshi2017towards}
\bibfield{author}{\bibinfo{person}{Finale Doshi-Velez} {and}
  \bibinfo{person}{Been Kim}.} \bibinfo{year}{2017}\natexlab{}.
\newblock \showarticletitle{Towards a rigorous science of interpretable machine
  learning}.
\newblock \bibinfo{journal}{\emph{arXiv preprint arXiv:1702.08608}}
  (\bibinfo{year}{2017}).
\newblock


\bibitem[\protect\citeauthoryear{Dove, Halskov, Forlizzi, and Zimmerman}{Dove
  et~al\mbox{.}}{2017}]%
        {dove2017ux}
\bibfield{author}{\bibinfo{person}{Graham Dove}, \bibinfo{person}{Kim Halskov},
  \bibinfo{person}{Jodi Forlizzi}, {and} \bibinfo{person}{John Zimmerman}.}
  \bibinfo{year}{2017}\natexlab{}.
\newblock \showarticletitle{UX design innovation: Challenges for working with
  machine learning as a design material}. In
  \bibinfo{booktitle}{\emph{Proceedings of the 2017 chi conference on human
  factors in computing systems}}. \bibinfo{pages}{278--288}.
\newblock


\bibitem[\protect\citeauthoryear{Eiband, Schneider, Bilandzic, Fazekas-Con,
  Haug, and Hussmann}{Eiband et~al\mbox{.}}{2018}]%
        {eiband2018bringing}
\bibfield{author}{\bibinfo{person}{Malin Eiband}, \bibinfo{person}{Hanna
  Schneider}, \bibinfo{person}{Mark Bilandzic}, \bibinfo{person}{Julian
  Fazekas-Con}, \bibinfo{person}{Mareike Haug}, {and} \bibinfo{person}{Heinrich
  Hussmann}.} \bibinfo{year}{2018}\natexlab{}.
\newblock \showarticletitle{Bringing transparency design into practice}. In
  \bibinfo{booktitle}{\emph{23rd international conference on intelligent user
  interfaces}}. \bibinfo{pages}{211--223}.
\newblock


\bibitem[\protect\citeauthoryear{Gebru, Morgenstern, Vecchione, Vaughan,
  Wallach, Daum{\'e}~III, and Crawford}{Gebru et~al\mbox{.}}{2018}]%
        {gebru2018datasheets}
\bibfield{author}{\bibinfo{person}{Timnit Gebru}, \bibinfo{person}{Jamie
  Morgenstern}, \bibinfo{person}{Briana Vecchione},
  \bibinfo{person}{Jennifer~Wortman Vaughan}, \bibinfo{person}{Hanna Wallach},
  \bibinfo{person}{Hal Daum{\'e}~III}, {and} \bibinfo{person}{Kate Crawford}.}
  \bibinfo{year}{2018}\natexlab{}.
\newblock \showarticletitle{Datasheets for datasets}.
\newblock \bibinfo{journal}{\emph{arXiv preprint arXiv:1803.09010}}
  (\bibinfo{year}{2018}).
\newblock


\bibitem[\protect\citeauthoryear{Ghai, Liao, Zhang, Bellamy, and Mueller}{Ghai
  et~al\mbox{.}}{2021}]%
        {ghai2021explainable}
\bibfield{author}{\bibinfo{person}{Bhavya Ghai}, \bibinfo{person}{Q~Vera Liao},
  \bibinfo{person}{Yunfeng Zhang}, \bibinfo{person}{Rachel Bellamy}, {and}
  \bibinfo{person}{Klaus Mueller}.} \bibinfo{year}{2021}\natexlab{}.
\newblock \showarticletitle{Explainable Active Learning (XAL) Toward AI
  Explanations as Interfaces for Machine Teachers}.
\newblock \bibinfo{journal}{\emph{Proceedings of the ACM on Human-Computer
  Interaction}} \bibinfo{volume}{4}, \bibinfo{number}{CSCW3}
  (\bibinfo{year}{2021}), \bibinfo{pages}{1--28}.
\newblock


\bibitem[\protect\citeauthoryear{Girardin and Lathia}{Girardin and
  Lathia}{2017}]%
        {girardin2017user}
\bibfield{author}{\bibinfo{person}{Fabien Girardin} {and} \bibinfo{person}{Neal
  Lathia}.} \bibinfo{year}{2017}\natexlab{}.
\newblock \showarticletitle{When User Experience Designers Partner with Data
  Scientists.}. In \bibinfo{booktitle}{\emph{AAAI Spring Symposia}}.
\newblock


\bibitem[\protect\citeauthoryear{Guidotti, Monreale, Ruggieri, Turini,
  Giannotti, and Pedreschi}{Guidotti et~al\mbox{.}}{2018}]%
        {guidotti2018survey}
\bibfield{author}{\bibinfo{person}{Riccardo Guidotti}, \bibinfo{person}{Anna
  Monreale}, \bibinfo{person}{Salvatore Ruggieri}, \bibinfo{person}{Franco
  Turini}, \bibinfo{person}{Fosca Giannotti}, {and} \bibinfo{person}{Dino
  Pedreschi}.} \bibinfo{year}{2018}\natexlab{}.
\newblock \showarticletitle{A survey of methods for explaining black box
  models}.
\newblock \bibinfo{journal}{\emph{ACM computing surveys (CSUR)}}
  \bibinfo{volume}{51}, \bibinfo{number}{5} (\bibinfo{year}{2018}),
  \bibinfo{pages}{1--42}.
\newblock


\bibitem[\protect\citeauthoryear{Gunning}{Gunning}{2017}]%
        {gunning2017explainable}
\bibfield{author}{\bibinfo{person}{David Gunning}.}
  \bibinfo{year}{2017}\natexlab{}.
\newblock \showarticletitle{Explainable artificial intelligence (xai)}.
\newblock \bibinfo{journal}{\emph{Defense Advanced Research Projects Agency
  (DARPA), nd Web}} \bibinfo{volume}{2}, \bibinfo{number}{2}
  (\bibinfo{year}{2017}).
\newblock


\bibitem[\protect\citeauthoryear{Hilton}{Hilton}{1990}]%
        {hilton1990conversational}
\bibfield{author}{\bibinfo{person}{Denis~J Hilton}.}
  \bibinfo{year}{1990}\natexlab{}.
\newblock \showarticletitle{Conversational processes and causal explanation.}
\newblock \bibinfo{journal}{\emph{Psychological Bulletin}}
  \bibinfo{volume}{107}, \bibinfo{number}{1} (\bibinfo{year}{1990}),
  \bibinfo{pages}{65}.
\newblock


\bibitem[\protect\citeauthoryear{Hind}{Hind}{2019}]%
        {hind2019explaining}
\bibfield{author}{\bibinfo{person}{Michael Hind}.}
  \bibinfo{year}{2019}\natexlab{}.
\newblock \showarticletitle{Explaining explainable AI}.
\newblock \bibinfo{journal}{\emph{XRDS: Crossroads, The ACM Magazine for
  Students}} \bibinfo{volume}{25}, \bibinfo{number}{3} (\bibinfo{year}{2019}),
  \bibinfo{pages}{16--19}.
\newblock


\bibitem[\protect\citeauthoryear{Hohman, Head, Caruana, DeLine, and
  Drucker}{Hohman et~al\mbox{.}}{2019}]%
        {hohman2019gamut}
\bibfield{author}{\bibinfo{person}{Fred Hohman}, \bibinfo{person}{Andrew Head},
  \bibinfo{person}{Rich Caruana}, \bibinfo{person}{Robert DeLine}, {and}
  \bibinfo{person}{Steven~M Drucker}.} \bibinfo{year}{2019}\natexlab{}.
\newblock \showarticletitle{Gamut: A design probe to understand how data
  scientists understand machine learning models}. In
  \bibinfo{booktitle}{\emph{Proceedings of the 2019 CHI conference on human
  factors in computing systems}}. \bibinfo{pages}{1--13}.
\newblock


\bibitem[\protect\citeauthoryear{Kaur, Nori, Jenkins, Caruana, Wallach, and
  Wortman~Vaughan}{Kaur et~al\mbox{.}}{2020}]%
        {kaur2020interpreting}
\bibfield{author}{\bibinfo{person}{Harmanpreet Kaur}, \bibinfo{person}{Harsha
  Nori}, \bibinfo{person}{Samuel Jenkins}, \bibinfo{person}{Rich Caruana},
  \bibinfo{person}{Hanna Wallach}, {and} \bibinfo{person}{Jennifer
  Wortman~Vaughan}.} \bibinfo{year}{2020}\natexlab{}.
\newblock \showarticletitle{Interpreting Interpretability: Understanding Data
  Scientists' Use of Interpretability Tools for Machine Learning}. In
  \bibinfo{booktitle}{\emph{Proceedings of the 2020 CHI Conference on Human
  Factors in Computing Systems}}. \bibinfo{pages}{1--14}.
\newblock


\bibitem[\protect\citeauthoryear{Kun, Mulder, De~G{\"o}tzen, and Kortuem}{Kun
  et~al\mbox{.}}{2019}]%
        {kun2019creative}
\bibfield{author}{\bibinfo{person}{Peter Kun}, \bibinfo{person}{Ingrid Mulder},
  \bibinfo{person}{Amalia De~G{\"o}tzen}, {and} \bibinfo{person}{Gerd
  Kortuem}.} \bibinfo{year}{2019}\natexlab{}.
\newblock \showarticletitle{Creative data work in the design process}.
\newblock In \bibinfo{booktitle}{\emph{Proceedings of the 2019 on Creativity
  and Cognition}}. \bibinfo{pages}{346--358}.
\newblock


\bibitem[\protect\citeauthoryear{Lai, Liu, and Tan}{Lai et~al\mbox{.}}{2020}]%
        {lai2020chicago}
\bibfield{author}{\bibinfo{person}{Vivian Lai}, \bibinfo{person}{Han Liu},
  {and} \bibinfo{person}{Chenhao Tan}.} \bibinfo{year}{2020}\natexlab{}.
\newblock \showarticletitle{" Why is' Chicago'deceptive?" Towards Building
  Model-Driven Tutorials for Humans}. In \bibinfo{booktitle}{\emph{Proceedings
  of the 2020 CHI Conference on Human Factors in Computing Systems}}.
  \bibinfo{pages}{1--13}.
\newblock


\bibitem[\protect\citeauthoryear{Lee, Siewiorek, Smailagic, Bernardino, and
  Berm{\'u}dez~i Badia}{Lee et~al\mbox{.}}{2020}]%
        {lee2020co}
\bibfield{author}{\bibinfo{person}{Min~Hun Lee}, \bibinfo{person}{Daniel~P
  Siewiorek}, \bibinfo{person}{Asim Smailagic}, \bibinfo{person}{Alexandre
  Bernardino}, {and} \bibinfo{person}{Sergi Berm{\'u}dez~i Badia}.}
  \bibinfo{year}{2020}\natexlab{}.
\newblock \showarticletitle{Co-Design and Evaluation of an Intelligent Decision
  Support System for Stroke Rehabilitation Assessment}.
\newblock \bibinfo{journal}{\emph{Proceedings of the ACM on Human-Computer
  Interaction}} \bibinfo{volume}{4}, \bibinfo{number}{CSCW2}
  (\bibinfo{year}{2020}), \bibinfo{pages}{1--27}.
\newblock


\bibitem[\protect\citeauthoryear{Liao, Gruen, and Miller}{Liao
  et~al\mbox{.}}{2020}]%
        {liao2020questioning}
\bibfield{author}{\bibinfo{person}{Q~Vera Liao}, \bibinfo{person}{Daniel
  Gruen}, {and} \bibinfo{person}{Sarah Miller}.}
  \bibinfo{year}{2020}\natexlab{}.
\newblock \showarticletitle{Questioning the AI: informing design practices for
  explainable AI user experiences}. In \bibinfo{booktitle}{\emph{Proceedings of
  the 2020 CHI Conference on Human Factors in Computing Systems}}.
  \bibinfo{pages}{1--15}.
\newblock


\bibitem[\protect\citeauthoryear{Lim, Dey, and Avrahami}{Lim
  et~al\mbox{.}}{2009}]%
        {lim2009and}
\bibfield{author}{\bibinfo{person}{Brian~Y Lim}, \bibinfo{person}{Anind~K Dey},
  {and} \bibinfo{person}{Daniel Avrahami}.} \bibinfo{year}{2009}\natexlab{}.
\newblock \showarticletitle{Why and why not explanations improve the
  intelligibility of context-aware intelligent systems}. In
  \bibinfo{booktitle}{\emph{Proceedings of the SIGCHI Conference on Human
  Factors in Computing Systems}}. \bibinfo{pages}{2119--2128}.
\newblock


\bibitem[\protect\citeauthoryear{Lipton}{Lipton}{2018}]%
        {lipton2018mythos}
\bibfield{author}{\bibinfo{person}{Zachary~C Lipton}.}
  \bibinfo{year}{2018}\natexlab{}.
\newblock \showarticletitle{The Mythos of Model Interpretability: In machine
  learning, the concept of interpretability is both important and slippery.}
\newblock \bibinfo{journal}{\emph{Queue}} \bibinfo{volume}{16},
  \bibinfo{number}{3} (\bibinfo{year}{2018}), \bibinfo{pages}{31--57}.
\newblock


\bibitem[\protect\citeauthoryear{Lundberg and Lee}{Lundberg and Lee}{2017}]%
        {lundberg2017unified}
\bibfield{author}{\bibinfo{person}{Scott Lundberg} {and} \bibinfo{person}{Su-In
  Lee}.} \bibinfo{year}{2017}\natexlab{}.
\newblock \showarticletitle{A unified approach to interpreting model
  predictions}.
\newblock \bibinfo{journal}{\emph{arXiv preprint arXiv:1705.07874}}
  (\bibinfo{year}{2017}).
\newblock


\bibitem[\protect\citeauthoryear{Madaio, Stark, Wortman~Vaughan, and
  Wallach}{Madaio et~al\mbox{.}}{2020}]%
        {madaio2020co}
\bibfield{author}{\bibinfo{person}{Michael~A Madaio}, \bibinfo{person}{Luke
  Stark}, \bibinfo{person}{Jennifer Wortman~Vaughan}, {and}
  \bibinfo{person}{Hanna Wallach}.} \bibinfo{year}{2020}\natexlab{}.
\newblock \showarticletitle{Co-designing checklists to understand
  organizational challenges and opportunities around fairness in ai}. In
  \bibinfo{booktitle}{\emph{Proceedings of the 2020 CHI Conference on Human
  Factors in Computing Systems}}. \bibinfo{pages}{1--14}.
\newblock


\bibitem[\protect\citeauthoryear{Miller}{Miller}{2019}]%
        {miller2019explanation}
\bibfield{author}{\bibinfo{person}{Tim Miller}.}
  \bibinfo{year}{2019}\natexlab{}.
\newblock \showarticletitle{Explanation in artificial intelligence: Insights
  from the social sciences}.
\newblock \bibinfo{journal}{\emph{Artificial intelligence}}
  \bibinfo{volume}{267} (\bibinfo{year}{2019}), \bibinfo{pages}{1--38}.
\newblock


\bibitem[\protect\citeauthoryear{Mittelstadt, Russell, and Wachter}{Mittelstadt
  et~al\mbox{.}}{2019}]%
        {mittelstadt2019explaining}
\bibfield{author}{\bibinfo{person}{Brent Mittelstadt}, \bibinfo{person}{Chris
  Russell}, {and} \bibinfo{person}{Sandra Wachter}.}
  \bibinfo{year}{2019}\natexlab{}.
\newblock \showarticletitle{Explaining explanations in AI}. In
  \bibinfo{booktitle}{\emph{Proceedings of the conference on fairness,
  accountability, and transparency}}. \bibinfo{pages}{279--288}.
\newblock


\bibitem[\protect\citeauthoryear{Ribeiro, Singh, and Guestrin}{Ribeiro
  et~al\mbox{.}}{2016}]%
        {ribeiro2016should}
\bibfield{author}{\bibinfo{person}{Marco~Tulio Ribeiro},
  \bibinfo{person}{Sameer Singh}, {and} \bibinfo{person}{Carlos Guestrin}.}
  \bibinfo{year}{2016}\natexlab{}.
\newblock \showarticletitle{" Why should i trust you?" Explaining the
  predictions of any classifier}. In \bibinfo{booktitle}{\emph{Proceedings of
  the 22nd ACM SIGKDD international conference on knowledge discovery and data
  mining}}. \bibinfo{pages}{1135--1144}.
\newblock


\bibitem[\protect\citeauthoryear{Ribeiro, Singh, and Guestrin}{Ribeiro
  et~al\mbox{.}}{2018}]%
        {ribeiro2018anchors}
\bibfield{author}{\bibinfo{person}{Marco~Tulio Ribeiro},
  \bibinfo{person}{Sameer Singh}, {and} \bibinfo{person}{Carlos Guestrin}.}
  \bibinfo{year}{2018}\natexlab{}.
\newblock \showarticletitle{Anchors: High-precision model-agnostic
  explanations}. In \bibinfo{booktitle}{\emph{Proceedings of the AAAI
  Conference on Artificial Intelligence}}, Vol.~\bibinfo{volume}{32}.
\newblock


\bibitem[\protect\citeauthoryear{Rosenfeld and Richardson}{Rosenfeld and
  Richardson}{2019}]%
        {rosenfeld2019explainability}
\bibfield{author}{\bibinfo{person}{Avi Rosenfeld} {and}
  \bibinfo{person}{Ariella Richardson}.} \bibinfo{year}{2019}\natexlab{}.
\newblock \showarticletitle{Explainability in human--agent systems}.
\newblock \bibinfo{journal}{\emph{Autonomous Agents and Multi-Agent Systems}}
  \bibinfo{volume}{33}, \bibinfo{number}{6} (\bibinfo{year}{2019}),
  \bibinfo{pages}{673--705}.
\newblock


\bibitem[\protect\citeauthoryear{Rudin}{Rudin}{2019}]%
        {rudin2019stop}
\bibfield{author}{\bibinfo{person}{Cynthia Rudin}.}
  \bibinfo{year}{2019}\natexlab{}.
\newblock \showarticletitle{Stop explaining black box machine learning models
  for high stakes decisions and use interpretable models instead}.
\newblock \bibinfo{journal}{\emph{Nature Machine Intelligence}}
  \bibinfo{volume}{1}, \bibinfo{number}{5} (\bibinfo{year}{2019}),
  \bibinfo{pages}{206--215}.
\newblock


\bibitem[\protect\citeauthoryear{Sun, Zhou, Wu, Zhang, Zhang, and Xiang}{Sun
  et~al\mbox{.}}{2020}]%
        {sun2020developing}
\bibfield{author}{\bibinfo{person}{Lingyun Sun}, \bibinfo{person}{Zhibin Zhou},
  \bibinfo{person}{Wenqi Wu}, \bibinfo{person}{Yuyang Zhang},
  \bibinfo{person}{Rui Zhang}, {and} \bibinfo{person}{Wei Xiang}.}
  \bibinfo{year}{2020}\natexlab{}.
\newblock \showarticletitle{Developing a Toolkit for Prototyping Machine
  Learning-Empowered Products: The Design and Evaluation of ML-Rapid}.
\newblock \bibinfo{journal}{\emph{International Journal of Design}}
  \bibinfo{volume}{14}, \bibinfo{number}{2} (\bibinfo{year}{2020}),
  \bibinfo{pages}{35}.
\newblock


\bibitem[\protect\citeauthoryear{Swartout}{Swartout}{1983}]%
        {swartout1983xplain}
\bibfield{author}{\bibinfo{person}{William~R Swartout}.}
  \bibinfo{year}{1983}\natexlab{}.
\newblock \showarticletitle{XPLAIN: A system for creating and explaining expert
  consulting programs}.
\newblock \bibinfo{journal}{\emph{Artificial intelligence}}
  \bibinfo{volume}{21}, \bibinfo{number}{3} (\bibinfo{year}{1983}),
  \bibinfo{pages}{285--325}.
\newblock


\bibitem[\protect\citeauthoryear{Tomsett, Braines, Harborne, Preece, and
  Chakraborty}{Tomsett et~al\mbox{.}}{2018}]%
        {tomsett2018interpretable}
\bibfield{author}{\bibinfo{person}{Richard Tomsett}, \bibinfo{person}{Dave
  Braines}, \bibinfo{person}{Dan Harborne}, \bibinfo{person}{Alun Preece},
  {and} \bibinfo{person}{Supriyo Chakraborty}.}
  \bibinfo{year}{2018}\natexlab{}.
\newblock \showarticletitle{Interpretable to whom? A role-based model for
  analyzing interpretable machine learning systems}.
\newblock \bibinfo{journal}{\emph{arXiv preprint arXiv:1806.07552}}
  (\bibinfo{year}{2018}).
\newblock


\bibitem[\protect\citeauthoryear{van Allen}{van Allen}{2018}]%
        {van2018prototyping}
\bibfield{author}{\bibinfo{person}{Philip van Allen}.}
  \bibinfo{year}{2018}\natexlab{}.
\newblock \showarticletitle{Prototyping ways of prototyping AI}.
\newblock \bibinfo{journal}{\emph{Interactions}} \bibinfo{volume}{25},
  \bibinfo{number}{6} (\bibinfo{year}{2018}), \bibinfo{pages}{46--51}.
\newblock


\bibitem[\protect\citeauthoryear{Wang, Yang, Abdul, and Lim}{Wang
  et~al\mbox{.}}{2019}]%
        {wang2019designing}
\bibfield{author}{\bibinfo{person}{Danding Wang}, \bibinfo{person}{Qian Yang},
  \bibinfo{person}{Ashraf Abdul}, {and} \bibinfo{person}{Brian~Y Lim}.}
  \bibinfo{year}{2019}\natexlab{}.
\newblock \showarticletitle{Designing theory-driven user-centric explainable
  AI}. In \bibinfo{booktitle}{\emph{Proceedings of the 2019 CHI conference on
  human factors in computing systems}}. \bibinfo{pages}{1--15}.
\newblock


\bibitem[\protect\citeauthoryear{Wolf}{Wolf}{2019}]%
        {wolf2019explainability}
\bibfield{author}{\bibinfo{person}{Christine~T Wolf}.}
  \bibinfo{year}{2019}\natexlab{}.
\newblock \showarticletitle{Explainability scenarios: towards scenario-based
  XAI design}. In \bibinfo{booktitle}{\emph{Proceedings of the 24th
  International Conference on Intelligent User Interfaces}}.
  \bibinfo{pages}{252--257}.
\newblock


\bibitem[\protect\citeauthoryear{Xie, Chen, Kao, Gao, and Chen}{Xie
  et~al\mbox{.}}{2020}]%
        {xie2020chexplain}
\bibfield{author}{\bibinfo{person}{Yao Xie}, \bibinfo{person}{Melody Chen},
  \bibinfo{person}{David Kao}, \bibinfo{person}{Ge Gao}, {and}
  \bibinfo{person}{Xiang'Anthony' Chen}.} \bibinfo{year}{2020}\natexlab{}.
\newblock \showarticletitle{CheXplain: Enabling Physicians to Explore and
  Understand Data-Driven, AI-Enabled Medical Imaging Analysis}. In
  \bibinfo{booktitle}{\emph{Proceedings of the 2020 CHI Conference on Human
  Factors in Computing Systems}}. \bibinfo{pages}{1--13}.
\newblock


\bibitem[\protect\citeauthoryear{Yang, Huang, Scholtz, and Arendt}{Yang
  et~al\mbox{.}}{2020}]%
        {yang2020visual}
\bibfield{author}{\bibinfo{person}{Fumeng Yang}, \bibinfo{person}{Zhuanyi
  Huang}, \bibinfo{person}{Jean Scholtz}, {and} \bibinfo{person}{Dustin~L
  Arendt}.} \bibinfo{year}{2020}\natexlab{}.
\newblock \showarticletitle{How do visual explanations foster end users'
  appropriate trust in machine learning?}. In
  \bibinfo{booktitle}{\emph{Proceedings of the 25th International Conference on
  Intelligent User Interfaces}}. \bibinfo{pages}{189--201}.
\newblock


\bibitem[\protect\citeauthoryear{Yang}{Yang}{2019}]%
        {yang2019profiling}
\bibfield{author}{\bibinfo{person}{Qian Yang}.}
  \bibinfo{year}{2019}\natexlab{}.
\newblock \emph{\bibinfo{title}{Profiling Artificial Intelligence as a Material
  for User Experience Design}}.
\newblock \bibinfo{thesistype}{Ph.D. Dissertation}.
\newblock


\bibitem[\protect\citeauthoryear{Yang, Scuito, Zimmerman, Forlizzi, and
  Steinfeld}{Yang et~al\mbox{.}}{2018}]%
        {yang2018investigating}
\bibfield{author}{\bibinfo{person}{Qian Yang}, \bibinfo{person}{Alex Scuito},
  \bibinfo{person}{John Zimmerman}, \bibinfo{person}{Jodi Forlizzi}, {and}
  \bibinfo{person}{Aaron Steinfeld}.} \bibinfo{year}{2018}\natexlab{}.
\newblock \showarticletitle{Investigating how experienced UX designers
  effectively work with machine learning}. In
  \bibinfo{booktitle}{\emph{Proceedings of the 2018 Designing Interactive
  Systems Conference}}. \bibinfo{pages}{585--596}.
\newblock


\bibitem[\protect\citeauthoryear{Yu, Yuan, Terveen, Wu, Forlizzi, and Zhu}{Yu
  et~al\mbox{.}}{2020}]%
        {yu2020keeping}
\bibfield{author}{\bibinfo{person}{Bowen Yu}, \bibinfo{person}{Ye Yuan},
  \bibinfo{person}{Loren Terveen}, \bibinfo{person}{Zhiwei~Steven Wu},
  \bibinfo{person}{Jodi Forlizzi}, {and} \bibinfo{person}{Haiyi Zhu}.}
  \bibinfo{year}{2020}\natexlab{}.
\newblock \showarticletitle{Keeping designers in the loop: Communicating
  inherent algorithmic trade-offs across multiple objectives}. In
  \bibinfo{booktitle}{\emph{Proceedings of the 2020 ACM Designing Interactive
  Systems Conference}}. \bibinfo{pages}{1245--1257}.
\newblock


\end{thebibliography}

\end{document}